\documentclass[12pt,preprint]{aastex}
\usepackage{amssymb}
\usepackage{epsfig}
\usepackage{amsmath}
\usepackage{lscape}
\usepackage{longtable}

\newcommand{\etal}{et~al.~} 
\newcommand{\Sersic}{S\'{e}rsic}

\newcommand{\kms}{\ifmmode\,{\rm km}\,{\rm s}^{-1}\else km$\,$s$^{-1}$\fi} 
\newcommand{\magarc}{\ifmmode {{{{\rm mag}~{\rm arcsec}}^{-2}}} 
             \else {{{mag}$~${arcsec}$^{-2}$}} 
             \fi} 
                                                                                 
\def\Equ#1{Eq.~(\ref{eq:#1})} 
\def\se#1{\S\ref{sec:#1}} 
\def\Fig#1{Fig.~\ref{#1}}

\def\be{\begin{equation}} 
\def\ee{\end{equation}} 
\def\ifm#1{\relax\ifmmode#1\else$\mathsurround=0pt #1$\fi} 
                                                                                  
\def \spose#1{\hbox to 0pt{#1\hss}} 
\def \lta{\mathrel{\spose{\lower 3pt\hbox{$\sim$}} 
     \raise 2.0pt\hbox{$<$}}} 
\def \gta{\mathrel{\spose{\lower 3pt\hbox{$\sim$}} 
     \raise 2.0pt\hbox{$>$}}} 
\def\ltsima{$\; \buildrel < \over \sim \;$} 
\def\lsim{\lower.5ex\hbox{\ltsima}} 
\def\gtsima{$\; \buildrel > \over \sim \;$} 
\def\gsim{\lower.5ex\hbox{\gtsima}}

\def\kms{\ifmmode\,{\rm km}\,{\rm s}^{-1}\else km$\,$s$^{-1}$\fi}

\def\c28 {C$_{28}$}

\def \V22{V_{2.2}}

\begin{document}

\title{A Survey of 286 Virgo Cluster Galaxies at Optical {\it griz} and Near-IR $H$-band: Surface Brightness Profiles and Bulge-Disk Decompositions}

\author{Michael McDonald} 
\affil{Department of Astronomy, University of Maryland, College Park, MD} 
\email{mcdonald@astro.umd.edu} 
 
\author{St\'{e}phane Courteau} 
\affil{Department of Physics, Engineering Physics and Astronomy, 
 Queen's University, Kingston, ON, Canada} 
\email{courteau@astro.queensu.ca} 

\author{R. Brent Tully}  
\affil{Institute for Astronomy, University of Hawaii,  
2680 Woodlawn Drive, Honolulu, HI} 
\email{tully@ifa.hawaii.edu} 
 
\and 
\author{Joel Roediger}
\affil{Department of Physics, Engineering Physics and Astronomy, 
 Queen's University, Kingston, ON, Canada} 
\email{jroediger@astro.queensu.ca}

\begin{abstract}
We present $g$,$r$,$i$,$z$ and H-band surface brightness profiles
and bulge-disk decompositions for a morphologically-broad sample of
286 Virgo cluster catalog (VCC) galaxies.  The H-band data come from a
variety of sources including our survey of 171 VCC galaxies at the UH
2.2-m, CFHT and UKIRT telescopes, and another 115 galaxies from the
Two-Micron All-Sky Survey (2MASS) and GOLDMine archives.  The optical
data for all 286 VCC galaxies were extracted from Sloan Digital Sky
Survey (SDSS) images.  The H-band and SDSS $griz$ data were analyzed
in a homogeneous manner using our own software, yielding a consistent
set of deep, multi-band surface brightness profiles for each galaxy.
Average surface brightness profiles per morphological bin were created
in order to characterize the variety of galaxy light profiles across
the Hubble sequence.  The 1D bulge-disk decomposition parameters, as
well as non-parametric galaxy measures, such as effective radius,
effective surface brightness and light concentration, are presented
for all 286 VCC galaxies in each of the five optical/near-infrared
wavebands.  The profile decompositions account for bulge and disk
components, spiral arms, nucleus and atmospheric blurring.  The Virgo
spiral galaxy bulges typically have a \Sersic\ index $n\sim1$, while
elliptical galaxies prefer $n\sim2$.  No galaxy spheroid requires
$n>3$. The light profiles for 70\% of the Virgo elliptical galaxies
reveal the presence of both a spheroid and disk component. A more
in-depth discussion of the structural parameter trends can be found in
McDonald \etal (2009b). The data provided here should serve as a base
for studies of galaxy structure and stellar populations in the cluster
environment.  The galaxy light profiles and bulge-disk decomposition
results are available at the Centre de Donn\'{e}es astronomiques de
Strasbourg (CDS; {\tt http://cds.u-strasbg.fr/}) and the author's own
website ({\tt http://www.astro.queensu.ca/virgo}).
\end{abstract}

\section{Introduction}\label{sec:intro}

The advent of large-area near-infrared (NIR) surveys such
as the Two-Micron All Sky Survey (Skrutskie \etal 2006; hereafter 
2MASS) and the UKIRT Infrared Deep Sky Survey (Lawrence \etal 2007; hereafter UKIDSS)
has forever improved our understanding of 
galaxy structure and evolution. The NIR data provide an unobscured view of galaxy structure, due to the relative insensitivity of NIR light to dust extinction, while probing the oldest and most representative stellar populations by mass. Unfortunately, the
aforementioned surveys, designed largely for the study of
bright infrared sources, use short exposure times (7.8 and 40 seconds
for 2MASS and UKIDSS, respectively), and suffer from the tremendous
brightness of the NIR sky (typically $\sim$ 3 orders of magnitude brighter than at $r$-band), leading to rather shallow surface brightness limits for galaxy studies when compared to similarly-designed optical surveys.

The stability of the optical night sky, especially at $r$- and $i$-bands,
allows for deep, accurate optical surface photometry with relatively short
integrations and careful data processing.  The Sloan Digital Sky Survey
(Adelman-McCarthy \etal 2008; hereafter SDSS) has revolutionized optical
astronomy, increasing the wealth of available astronomical data (both
photometric and spectroscopic) by orders of magnitude.  A development
made possible by the SDSS was the unraveling of bimodalities in various
galaxy properties such as color, star formation rate and clustering
(e.g., Strateva \etal 2003; Blanton \etal 2003; Kauffmann \etal 2003;
Baldry \etal 2004; Brinchmann \etal 2004; Balogh \etal 2004).
These and other studies have provided firm evidence that galaxies
can be separated into two distinct classes: blue, star-forming galaxies
found primarily in the field, and red, quiescent galaxies found
primarily in clusters.

Prior to these findings, Tully \& Verheijen (1997; hereafter TV97)
reported a different form of galaxy structural bimodality. This one, independent of the SDSS
colour bimodality, was based on a sample of 62 members of the Ursa Major (UMa) cluster of galaxies; TV97 found a bimodal distribution in the distribution of K$^{\prime}$-band (2.2 $\mu$m) surface brightness of galaxy disks.  McDonald \etal (2009a), 
who re-analyzed the same UMa data using more sophisticated
light profile decompositions to measure the disk structural parameters,
corroborated TV97's results.
Bell \& de Blok (2000) have however argued that the UMa
sample was too small for a statistically significant proof of
bimodality.  It is partly this unresolved issue, as well as the  
need for a complete NIR survey of a nearby galaxy cluster to study the (dust-free) structural parameters and stellar populations of cluster galaxies, that motivated our NIR survey of the Virgo cluster. 

Using the optical and NIR data presented in this paper, McDonald \etal (2009b)
performed 1D bulge-disk decompositions for 286 Virgo cluster galaxy surface brightness profiles and bolstered the evidence for a disk surface brightness bimodality.
Furthermore, it was found for galaxies of all morphologies, that the 
\emph{effective} surface brightness (defined as the surface brightness
measured at the half-light radius) distribution had three peaks, each 
defining: (i) high surface brightness, gas-poor galaxies, (ii) low surface
brightness, gas-poor galaxies and high surface brightness, gas-rich galaxies,
and (iii) low surface brightness, gas-rich galaxies. A more thorough discussion
of these, and other results from this survey, can be found in McDonald \etal (2009b).

This paper presents the deep, optical and NIR surface brightness profiles
for our sample of 286 Virgo cluster galaxies.  The paper outline is as
follows: we describe in \S2 the sample selection and the optical and
NIR data acquisition.  In \S3, we give an overview of our NIR data reduction
methods and we describe in \S4 the process of surface brightness profile
extraction for both the optical and NIR data.  We discuss in \S5
the quality of the NIR data and compare our data products with
those from 2MASS and SDSS. The full collection of surface brightness
profiles is presented in \S6, along with a brief analysis of the
average profile shape for various galaxy types.  Galaxy structural
parameters from our light profile decompositions and from non-parametric
galaxy measurements are presented in \S7.  We conclude and ponder future work in \S8.

In this paper, we assume a distance of 16.5 Mpc or \emph{m-M}=31.18 for
all Virgo cluster galaxies (Mei \etal 2007).  At that distance, 
$1\arcsec$ = 80~pc.

\section{Virgo Sample}\label{sec:sample}

The construction of an unbiased distribution of galaxy surface brightness
requires volume completeness, which can be most easily obtained for galaxies
in a cluster which all lie at a common distance.  Our sample is drawn from the
Virgo cluster catalogue of Binggeli \etal (1985; hereafter VCC).  
The VCC catalogue contains 2096 galaxies within an area of 
$\sim$140 deg$^2$ on the sky, centered on the galaxy M87 at 
$\alpha$$\sim$12$^h$25$^m$ and $\delta$$\sim$13$^{\circ}$. 
The VCC is asserted to be complete down to a limiting absolute 
magnitude of $M_B \sim -13$ and to contain many objects as faint 
as $M_B \sim -11$.  The full sample was reduced to a manageable size of 286 galaxies by performing a brightness cut ($M_B \leq -15.15$ mag), a spatial cut (to remove contamination from W, W', and M background groups) and a velocity cut ($V_{rad}<$ 3000 \kms). These cuts, which yield a sample covering a wide range of luminosities and morphologies, are discussed in further detail in McDonald \etal (2009b). 

\subsection{NIR Data}\label{sec:nir}
Deep $H$-band imaging for some VCC galaxies is already 
available from the 2MASS and GOLDMine\footnote{\tt{http://goldmine.mib.infn.it/}} (Gavazzi \etal 2003)
databases. $H$-band images from GOLDMine were kindly provided by G. Gavazzi, while
calibrated 2MASS galaxy images were extracted from the online
database.  Many of the 2MASS and GOLDMine images were not deep enough
for our purposes.  Whilst adequate for large HSB galaxies, the
high 2MASS surface brightness threshold (typically $\mu_H = 21$ mag arcsec$^{-2}$
-- Bell \etal 2003; Courteau \etal 2007; Kirby \etal 2008)
undermines the use of that database for deep extragalactic studies. 
Likewise, only a fraction of the available GOLDMine images were deep enough
to properly separate bulge and disk light components, or suffered from ghost images. Whenever possible, we attempted to salvage such images by either masking the ghosts or confining our analysis to a single chip, which were typically unaffected by contamination due to ghosts.
We secured $H$-band imaging for the remainder (187 galaxies) of
our sample with the detectors ULBCam (UH~2.2-m), WFCAM (UKIRT), and WIRCAM (CFHT) over the period 
2005-2009.  The existing and new observations are summarized in Table 1 of McDonald \etal (2009b).

The surface brightness profiles for the deep GOLDMine and 2MASS 
images were measured using the same techniques as our new NIR
images to ensure uniformity for the entire database. Further details
regarding the data reduction process and quality are given below.

\subsection{Optical Data}\label{sec:SDSS}

We have extracted calibrated $ugriz$ images from the 
SDSS for the 286 VCC galaxies in our sample.
Surface brightness profiles and total luminosities
were extracted for all of these galaxies 
in all five SDSS bands using procedures described in Courteau (1996) and McDonald \etal (2009b). 
$u$-band images were consistently 
shallower than those in the $griz$ bands and, thus, were discarded. 
Sky levels for background subtraction and the photometric 
zero-points for calibration were obtained from the SDSS 
image headers and the SDSS archives, respectively. 
However, we will see in \se{sky} that we ultimately
prefer our own estimates of the sky background for each
of the galaxy images.
The remainder of the profile extraction technique is 
identical to that used for the NIR photometry, 
as we describe below. 

\section{NIR Observations and Data Reduction}\label{sec:NIR}
Virgo cluster galaxies which lacked suitable NIR imaging
were observed with ULBCam\footnote{ULBCam, the Ultra Low Background
Camera, is a JWST prototype camera developed at the IfA by Don Hall
and his team. It combines 4 2048$\times$2048 arrays, of which we
used only the cleanest one.} on the 2.2-m UH telescope (Hall \etal 2004),
WIRCAM on the 3.6-m CFHT (Puget \etal 2004) and WFCAM on the 3.8-m UKIRT (Hambly \etal 2008) from April 2005 to May 2008 (see Table 1 of McDonald \etal 2009b).

For all three NIR imagers, we only used one of the available detectors
given the relatively small size of our targets. A single ULBCam 
detector has a $8\farcm5 \times 8\farcm5$ field of view (FOV)
with a pixel scale of $0\farcs{25}/$pixel, while WFCAM has a
$13\farcm{6} \times 13\farcm{6}$ FOV with a pixel scale
of $0\farcs{40}/$pixel, and WIRCAM has a 
$10\farcm{2} \times 10\farcm{2}$ FOV with a pixel scale
of $0\farcs{30}/$pixel. Altogether, 129, 35, and 27 galaxies
were observed at H-band with ULBCam, WIRCAM and WFCAM, respectively.

Given the rapid NIR sky fluctuations, we used a maximum single frame
exposure time of 40s with ULBCam, 15s with WIRCAM and 10s with WFCAM
to maximize sky flux, while keeping within the detector's linear regime.
A dithering script was used to minimize resampling of bad pixels. In order to
ensure sampling of each galaxy out to several effective radii, we first
pre-classified the galaxies by eye as high, medium, and low surface brightness. For these
three classes we aimed for total exposure times of 8, 16, and 24 minutes (on the 2.2-m), respectively. These different exposure times were motivated by our desire for a well-sampled disk in order to obtain a reliable bulge-disk decomposition.

Basic flat-fielding, stacking and bad pixel rejection procedures were
applied to the ULBCam data using the XVista astronomical software
package\footnote{{{\tt http://astronomy.nmsu.edu/holtz/xvista/index.html}}}.
Primary data reductions for the WIRCAM and WFCAM made use of observatory
pipelines. The removal of geometric distortion and photometric
calibration of the ULBCam data, as well as the procedures for
extraction of surface brightness profiles for all NIR and optical
data, are described below.

\subsection{ULBCam Geometric Distortion}
We corrected the significant geometric distortions in ULBCam images
by observing the dense star field FS17 and computing the offset between
the predicted and observed stellar positions, following Meurer \etal (2002).
 The following simple correction minimizes
 the distortion in our images:
 \[ \Delta x(pix)=16.64-0.0162y \]
 \[ \Delta y(pix)=0.0182x -0.0230y +3.80 \]

\begin{figure*}[htb]
\centering
\includegraphics[width=0.8\textwidth]{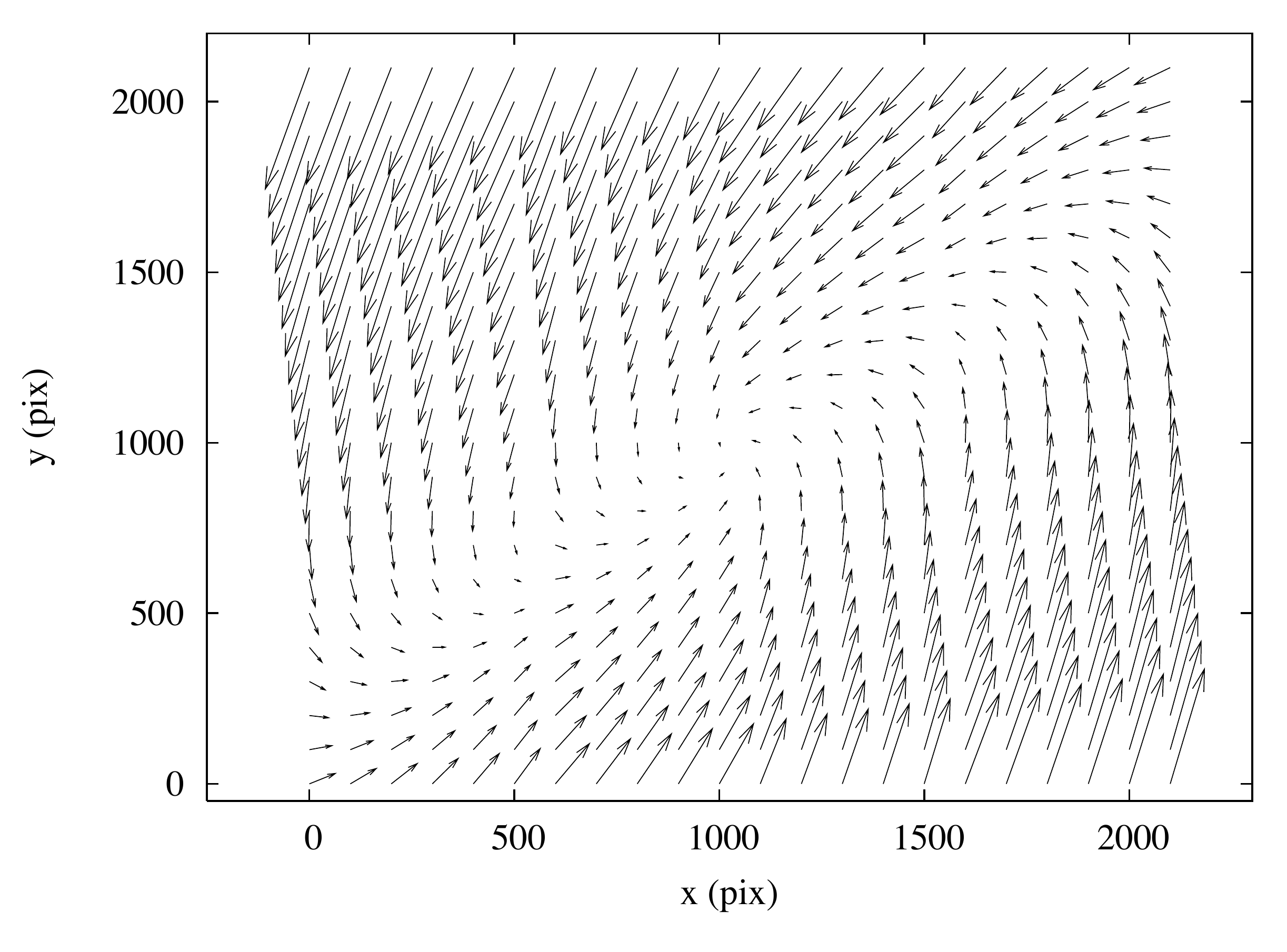}
\caption{Geometric distortion in one array of the ULBCam
  4096$\times$4096 detector. The magnitudes of the vectors have been
  increased by a factor of 8 for clarity.}
\label{distort}
\end{figure*}

Additionally, each ULBCam 2048x2048 detector is comprised of four smaller
512x2048 arrays that required horizontal offsets of -4 pixels (arrays 1 and 3) to account for 
their physical separation prior to determining the detector-wide 
astrometric correction.  
The 2D pattern of this correction model, shown 
in \Fig{distort}, is accurate to within 1 pixel (0.25\arcsec) across 
the full FOV.

\subsection{NIR Flux Calibration}\label{sec:2MASS}
\begin{figure*}[htb]
\centering
\includegraphics[width=0.8\textwidth]{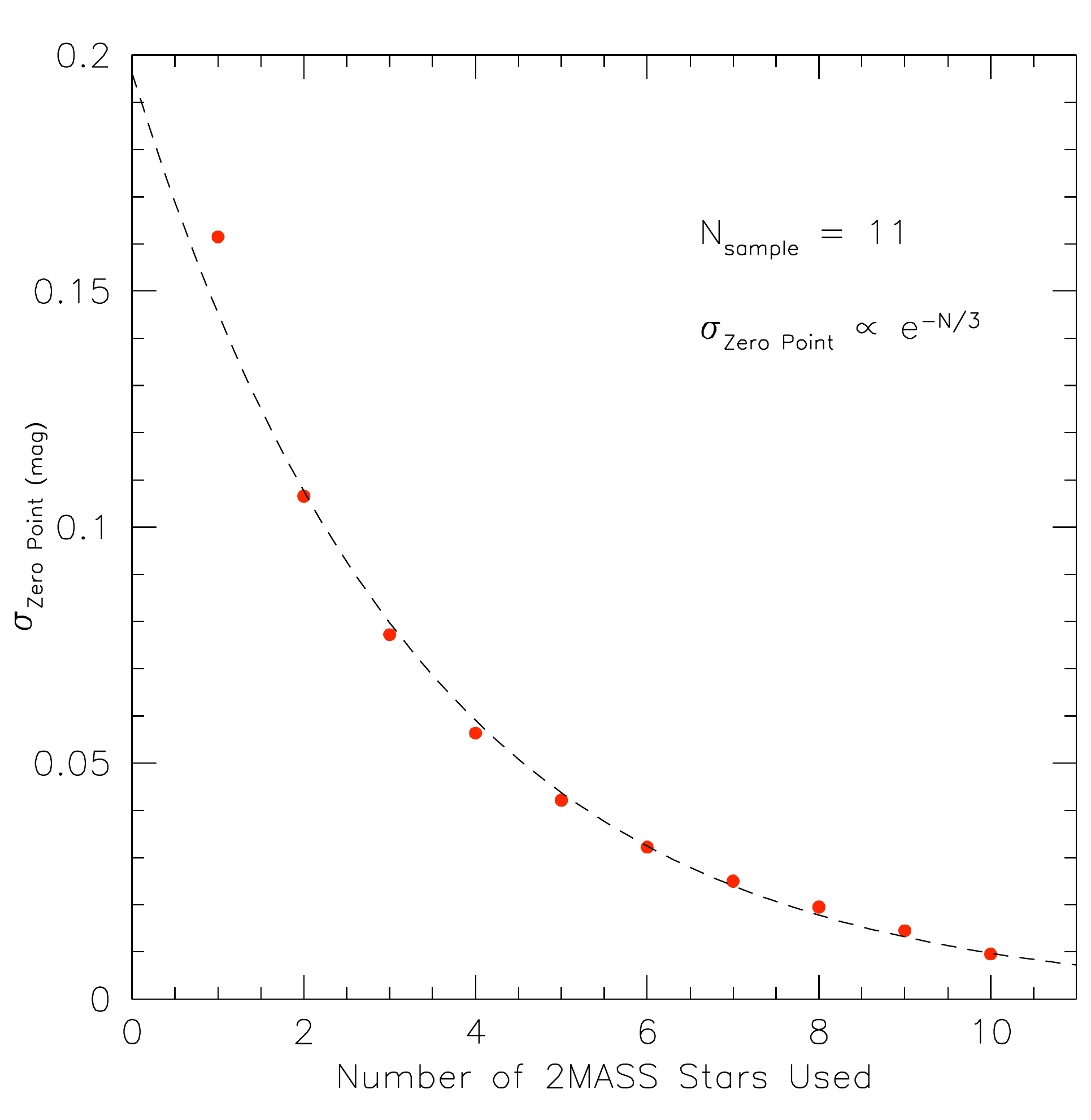}
\caption{Calibration error as a function of the number of 
field stars used per calibration. In order to compute the standard
deviation, $N$ stars were chosen at random and the mean brightness was
computed. This process was repeated 10,000 times to ensure every
possible combination of stars was used. The standard deviation represents the
difference in means depending on which $N$ out of 11 stars were chosen.}
\label{calib}
\end{figure*}

The target flux calibration is based on 2MASS foreground 
stars.  For each galaxy field, all stars above a given flux 
level (corresponding to the minimum detection level in 2MASS) 
were marked and their coordinates cross-correlated with the 
2MASS stellar library.  Typically, three to seven 2MASS stars per ULBCam 
galaxy field were marked.  The ratio of our instrumental magnitudes 
for those stars to those provided by 2MASS enabled a calibration
of all of our photometry at each pixel, independent of any
airmass or photometric variations (our NIR images were still
all obtained in most favorable conditions).  This approach was 
tested extensively against the traditional method of multiple 
standard star observations per night and found to be equally 
reliable provided that at least three 2MASS stars were used per field.
Under such conditions, the standard deviation of the derived 
zero-point corrections from each star is $\sim$0.1 mag, as 
seen in \Fig{calib}.  This error is comparable to those quoted 
for NIR photometric calibrations using standard star fields 
(de Jong 1996; Gavazzi \etal 1996; MacArthur \etal 2003).

\section{Surface Brightness Profile Extraction}
The measurements of galaxy surface brightnesses rely on the
careful mapping, in one or two dimensions, of a representative
light profile.  Our profile extraction uses numerous XVISTA
routines and procedures described in Courteau (1996).  We review
the important steps below.  While azimuthally-averaged surface
brightness profiles for some of our galaxies are available
elsewhere (e.g. 2MASS and GOLDMine), we have recovered original
images in all cases and rederived our own surface brightness
profiles for all of these galaxies, to ensure a homogeneous database.

\subsection{Sky Measurement}\label{sec:sky}

A sky level error of only $\pm$0.01\% at
NIR wavelengths can result in either an artificial truncation
(over-subtraction) or upturn (under-subtraction) of the galaxy
SB profile at large radii.  At H-band, this truncation/upturn would occur
at $\sim$ 23.5 mag arcsec$^{-2}$. An accurate assessment of the sky
level is thus crucial for the extraction of a reliable, deep, NIR
surface brightness profile.

For all images, the sky level is estimated in the final fully-reduced
image by isolating five regions (by eye) away from the galaxy, free of
any other sources, and calculating the mode of the sky intensities
per pixel within each sky region. The average and standard deviation of the five sky values were then computed. 
Sky levels were estimated this way not only for our data but for
2MASS and GOLDMine images as well. 
The typical sky fluctuation for our NIR data is $\sim$0.005\%. 

\begin{figure*}[htb]
\centering
\includegraphics[width=0.8\textwidth]{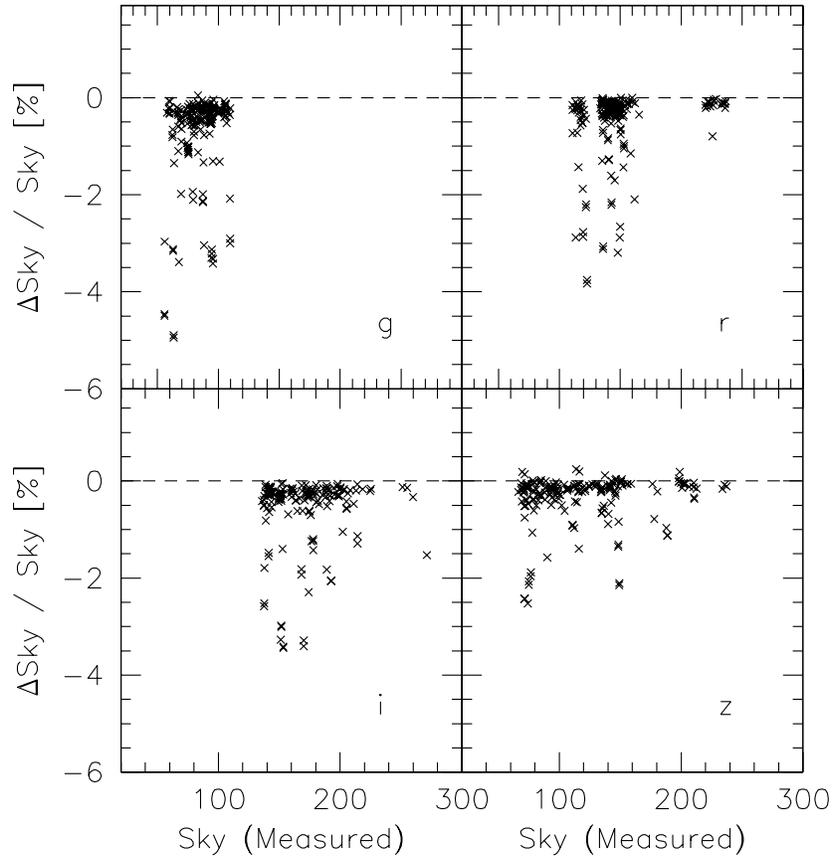}
\caption{Difference between our interactively-measured and 
  the SDSS-provided sky flux as a function of our measured
  sky level (in counts). Negative Y-values indicate a larger
  SDSS value.}
\label{skycompare}
\end{figure*}

Thanks to its extended sky coverage, SDSS should achieve accurate
sky level measurements
({{\tt http://www.sdss.org/dr5/algorithms/flat\_field.html}}).  
However, we still use our interactive assessment of the sky background,
as described above, for all SDSS data. Fig. \ref{skycompare} shows the
difference between our measured sky values and those provided by
SDSS. The latter is always biased high, likely due to the inclusion of bright, extended sources.  Our interactive technique
ensures a (mostly) contaminant-free selection of sky fields. However, due to the fact that we extend the surface brightness profiles to the edge of the field of view, the uncertainty in the profile shape is dominated by systematic (sky measurement) errors at large radii. While
seemingly small, sky level errors of a few percent, as seen in 
Fig. \ref{skycompare}, yield significant deviations in a galaxy
surface brightness profile (McDonald \etal 2009b). We find that the shape of the surface brightness profile in $g$-band is typically dominated by sky measurement errors by $\mu_g \sim $26 mag arcsec$^{-2}$. Below this surface brightness level (and corresponding surface brightnesses in $r$,$i$,$z$) users should consider the data with 1-sigma sky error envelopes.

\subsection{PSF Measurement and Star Masking}
The seeing point-spread functions (PSF) is often
modeled either as a Gaussian or a Moffat (1969) function. The latter has
broader wings, which better matches our stellar fields;  
we have thus used the Moffat function throughout our analysis.
On average, 17 stars per field were modeled for PSF measurements. 
The average FWHM for all ULBCam observations taken over 2005-2008 was $1.2 \pm 0.2^{\prime\prime}$; slightly better
imaging was achieved at CFHT and UKIRT.  The average FWHM
measurement for each image is later used to convolve the analytical
models for the galaxy bulge and disk light (see McDonald \etal 2009a).

The identification of foreground stars also enables their removal
from the galaxy light. The use of circular masks with radii equal to $4\times$
FWHM ensures that most of this contaminant light is properly removed.
Our automated masking routine could not identify bright stars with
diffraction spikes or other irregularly shaped features; those
obtrusive objects were manually masked.

\subsection{Isophotal Fitting}
Surface brightness profiles were extracted for all of our galaxy images. The XVISTA command, PROFILE,
performs this operation through isophotal fitting, using a generalized non-linear least-squares
fitting routine.  The ellipticity and position angle, but not the center, of each
elliptical isophote are allowed to vary. The center is determined by measuring the centroid of the central
light distribution 
with the XVISTA command AXES.  Beyond 1-2 disk scale lengths, the 
position angle and ellipticity usually settle to constant values. 
The isophotal solution based on these values is extrapolated to larger
radii where the signal-to-noise is too low for PROFILE to converge
on a unique solution. Adjustments to the ellipticity profile can 
be made with the XVISTA command, CPROF, to account for abrupt
changes due to any non-axisymmetric features. 
Ellipse fitting to map the radial surface brightness profile
requires the galaxy to be slightly (but not fully!) inclined,
so that the azimuthal direction is projected onto the plane of
the sky.  Edge-on galaxies are thus excluded from this procedure. 
Fortunately, only 3 galaxies in our sample have inclinations
$>80^\circ$; non-parametric structural measures (e.g., effective
surface brightness) for those galaxies are still valid. 

\subsection{Profile Depth and Signal-to-Noise Ratios}

We calculate the signal-to-noise (hereafter $S/N$) ratio
of a surface brightness profile as a function of radius as:

\begin{equation}
{\frac{S}{N}(r) = \frac{I_t(r) /pix^2 \times \sqrt{A_{ell}} pix}{\sqrt{I_{sky}} /pix}}
\label{eq:SNR}
\end{equation}
where the number of pixels (expressed as an area) along each isophote is given by: 
\begin{equation}
A_{ell} = 2\pi \sqrt{0.5(a^2 + b^2)}, 
\end{equation}

\noindent{}where $a$, $b$ are the semi-major and semi-minor axes of a given elliptical
isophote, $I_t(r)$ is its total surface brightness level (intensity in
counts), and $I_{sky}$ is the surface brightness of the sky.

The measurement of $I_{sky}$ was outlined in \se{sky}. According to
\Equ{SNR}, depths of $\sim 23.5~H$-\magarc with $S/N\sim$3 can be achieved
with an exposure time of 480s (at the 2.2-m UH telescope).
For intermediate and low surface brightness galaxies, our exposure
times of 960s and 1440s were determined to yield $S/N\sim$ 3 at
surface brightness depths of 24 and 24.5~$H$-\magarc, respectively.
These levels correspond to $\sim$4 disk scale lengths for disk
galaxies and $\sim$5 effective radii for spheroidal systems.

\section{NIR Data Quality}\label{sec:quality}

While our collection of Virgo cluster galaxy light profiles 
results from the merging of independent imaging surveys, we have
imposed our own uniform analysis methods to the entire database.
Internal errors can be assessed from repeat measurements, and
we can compare our extracted light profiles with those from
the original authors, whenever available.  We can also compare
our estimates of empirical quantities such as total luminosity,
scale radii and concentration to those provided with surveys
such as 2MASS and SDSS in order to test for systematic
differences.

\subsection{Independent Calibration Errors from Multiple Measurements}

\begin{figure*}[htb]
\centering
\includegraphics[width=0.8\textwidth]{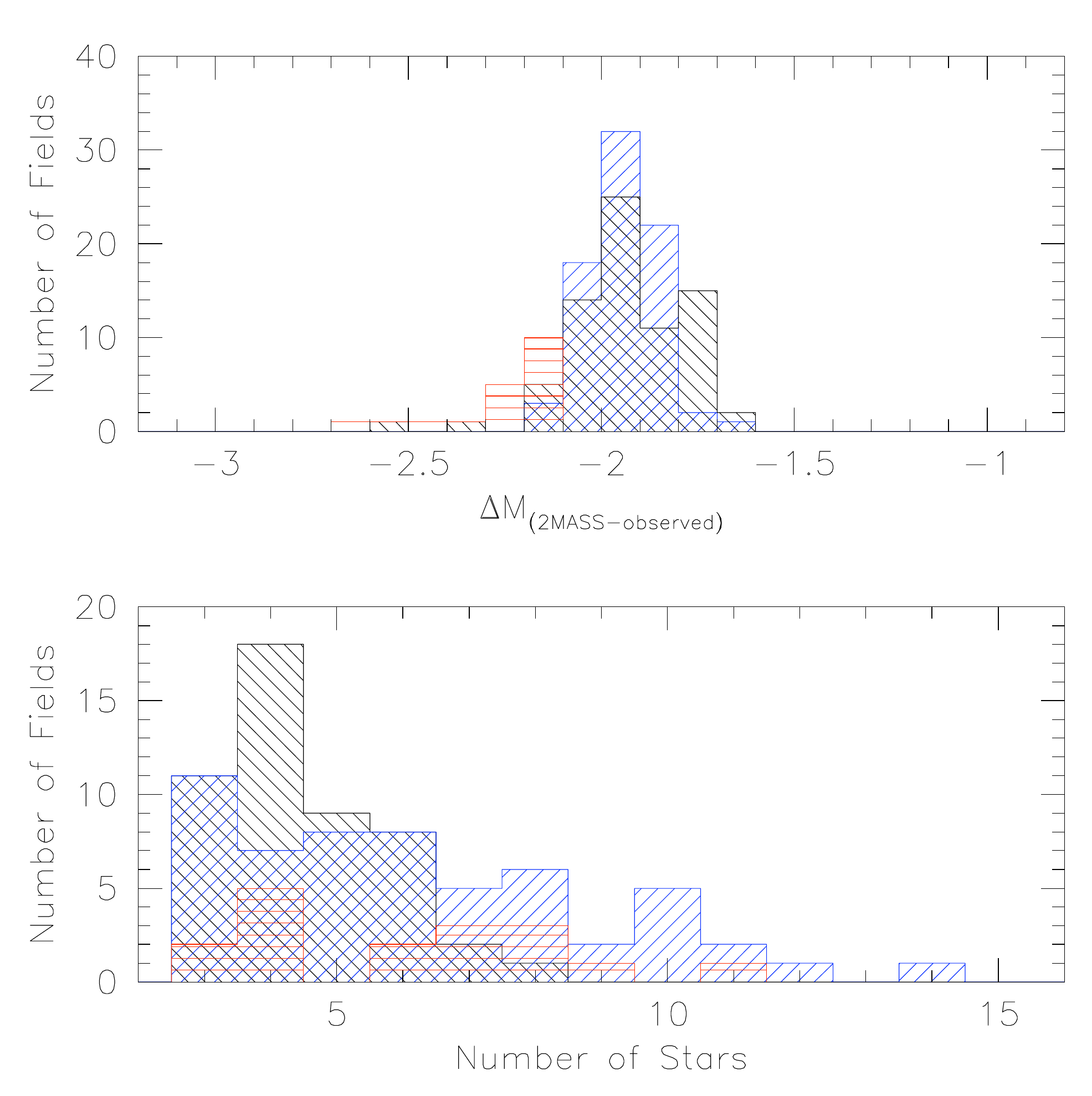} 
\caption{Upper panel: Distribution of magnitude offsets
  between observed and 2MASS measurements for foreground stars over
  many observing runs. The black, red and blue histograms correspond
  to observations taken in 2005, 2006 and 2007, respectively.  Lower
  panel: Number of foreground stars used to determine a relative
  magnitude offset per single observation; a minimum of 3 stars 
  per field was found to be necessary to constrain the photometric zeropoint.}
\label{zeropoint-cal}
\end{figure*}

We have repeat measurements for three galaxies observed 
from 2005 to 2007: VCC1614, VCC1516 and VCC1664. We find no variation
in the surface brightness profiles over these three epochs,  demonstrating the stability of our
calibrations over time.  Since our calibration of ULBCam
data is tied to the 2MASS system, we did not derive
extinction coefficients and zero-point corrections
for each observing night.  However, we can compute relative
offsets between the observed and 2MASS magnitudes for several
stars in each observed field.  The distribution of the
photometric offsets shown in \Fig{zeropoint-cal} spans 4 years of 
observations and a full range of airmasses. The measured width of
$\sim 0.2 H$ mag for this distribution is expected for night-to-night
zero-point variations alone (e.g., Courteau 1996).

\subsection{Profile Comparison with Independent Measurements}

\begin{figure*}[htb]
\centering
\includegraphics[width=0.8\textwidth]{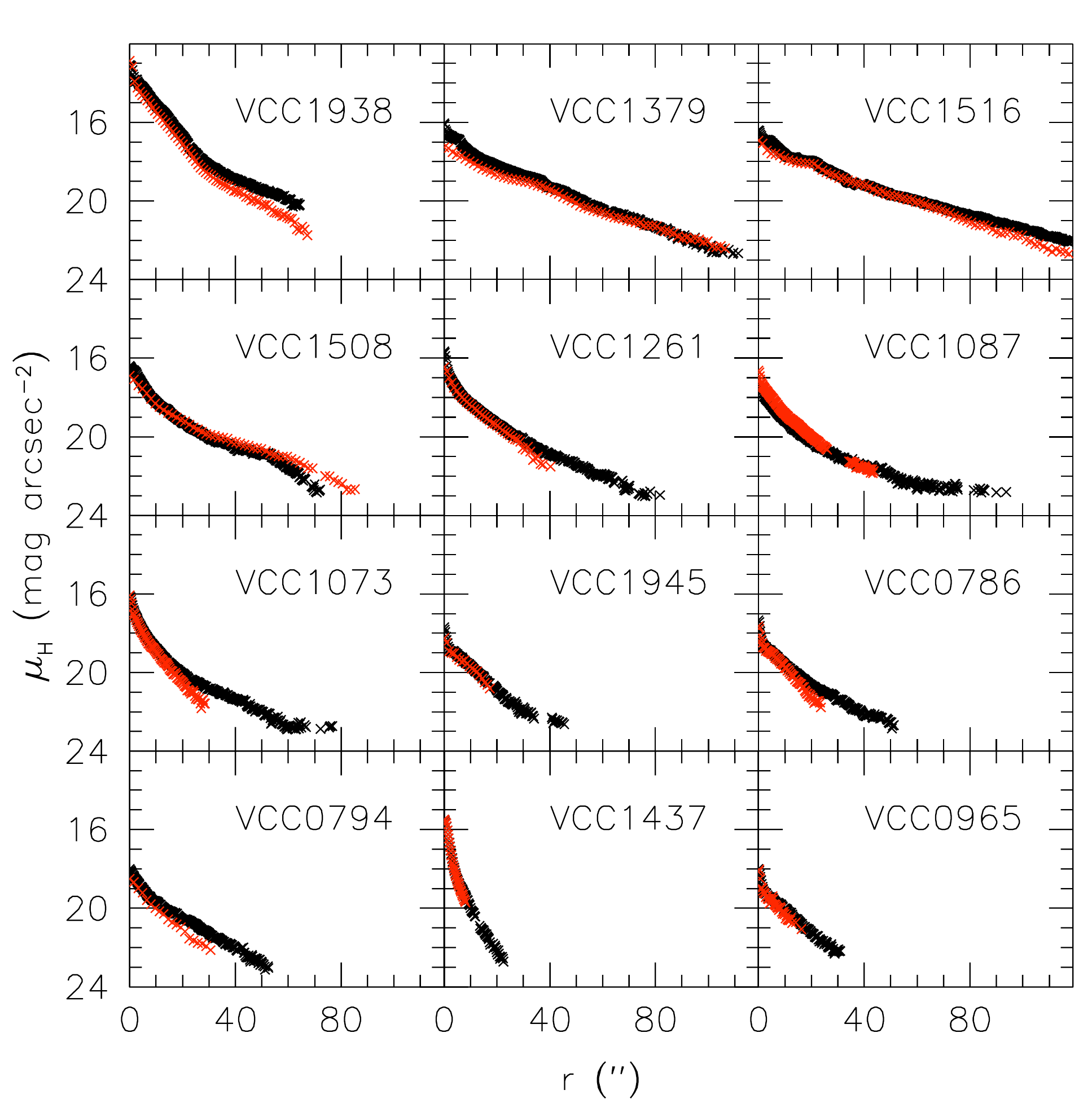} 
\caption{Comparison of GOLDMine (red) and ULBCam (black) $H$-band surface
  brightness profiles for a sample of matching galaxies, ordered (from left to right and top to bottom)
  by total luminosity.  The good match in profile depth found at the bright end of the sample is lost
  for fainter galaxies.}
\label{gavazzi-compare}
\end{figure*}

A total of 80 GOLDMine galaxies that satisfy
our $Q$ criterion (McDonald \etal 2009b) were included in our sample.  Original
images from GOLDMine's heterogeneous collection of galaxies were
kindly provided to us by G. Gavazzi.  The average $H$-band surface
brightness zero-point offset of $\pm$ 0.15 \magarc between our
ULBCam and GOLDMine's original light profiles is small enough
to justify merging the two samples into one.  However, for
complete uniformity, we have recomputed surface brightness
profiles for GOLDMine galaxies using our own data reduction
techniques outlined earlier.  Another motivation for doing so
is the desire to have linearly, rather than logarithmically (as in
GOLDMine), sampled surface brightness profiles.  This exercise
also led to more data points in the outer disks, which is relevant
for the extrapolation of light profiles to infinity.
\Fig{gavazzi-compare} shows the overlap of our profiles
derived from ULBCam and GOLDMine images for selection of VCC galaxies covering a broad range in luminosity and morphology.

\begin{figure*}[htb]
\centering
\includegraphics[width=0.8\textwidth]{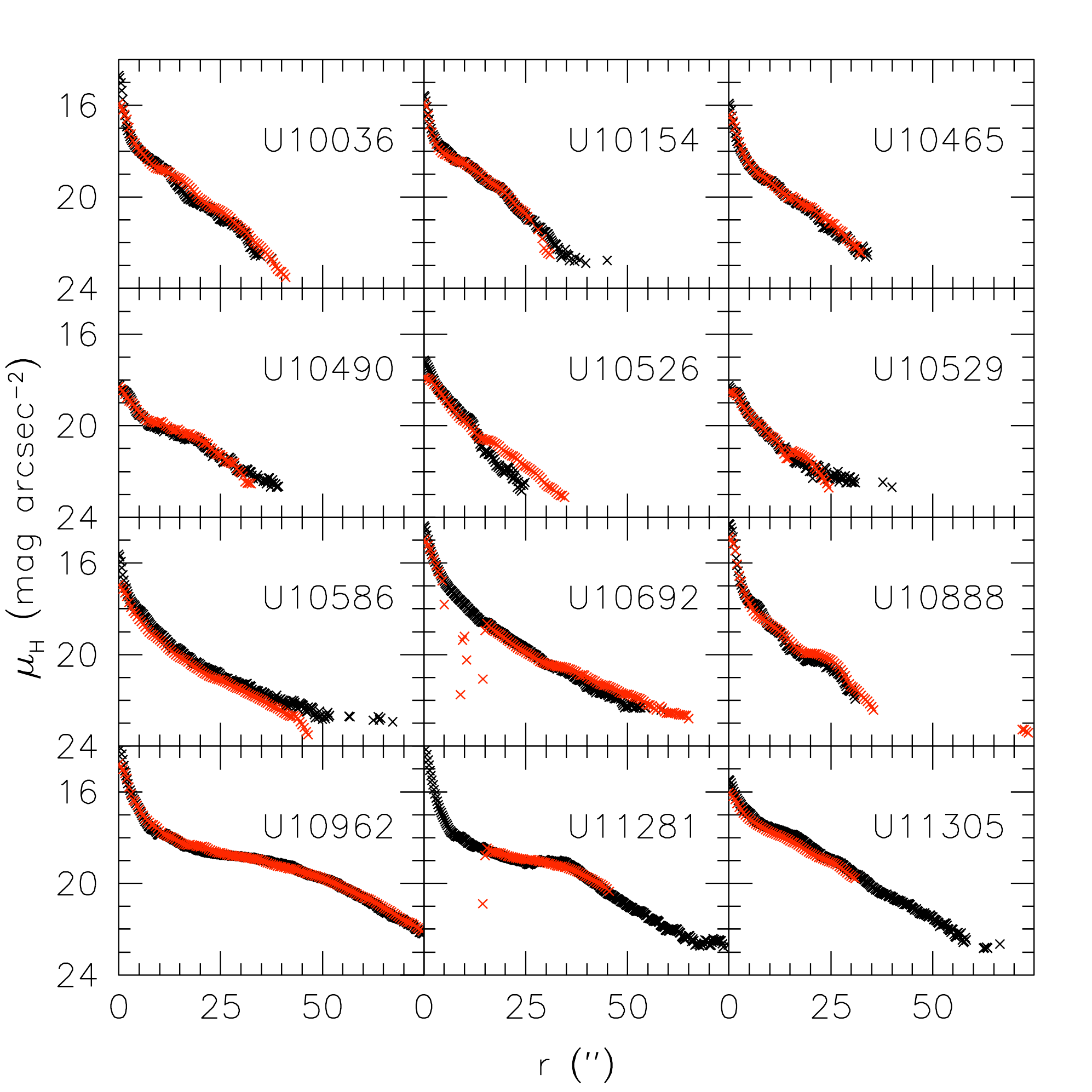} 
\caption{Comparison of M03 (red) and ULBCam (black) $H$-band surface
  brightness profiles. These galaxies, while not members of the Virgo cluster, were observed using ULBCam in order to provide an independent comparison of our calibration and analysis techniques.}
\label{dust-compare}
\end{figure*}

We now compare our calibrated profiles with the fully
homogeneous database of deep $H$-band images of late-type
northern spirals by MacArthur \etal (2003; hereafter M03).
These $H$-band images of spiral galaxies were collected at 
the KPNO 4-m telescope with the COB detector; typical total
integration times were 20 mins per galaxy.  We observed 12
non-Virgo UGC galaxies from M03's sample with ULBCam in
April 2007 for comparison.  As shown in \Fig{dust-compare},
the match with M03 is typically very good, with a mean deviation
between the two samples of $\sim 0.1$ $\magarc$. 
This comparison with an independent database reinforces
the quality of our dataset.

To summarize this section, we have shown that we can merge
our $Q>0.5$ GOLDMine and ULBCam $H$-band SB profiles into one
sample with a systematic error $<0.15$ $\magarc$.  Because
all profiles are calibrated to 2MASS, the 2MASS profiles
with $Q>0.5$ are also naturally integrated into our system.
Our entire collection of $H$-band SB profiles is thus self-consistent 
to within $<$~0.15 $\magarc$ out to the last measured data point.  

\subsection{Comparison of 2MASS and SDSS Data Products}\label{sec:other}

\begin{figure*}[htb]
\centering
\begin{tabular} {c}
\includegraphics[width=0.45\textwidth]{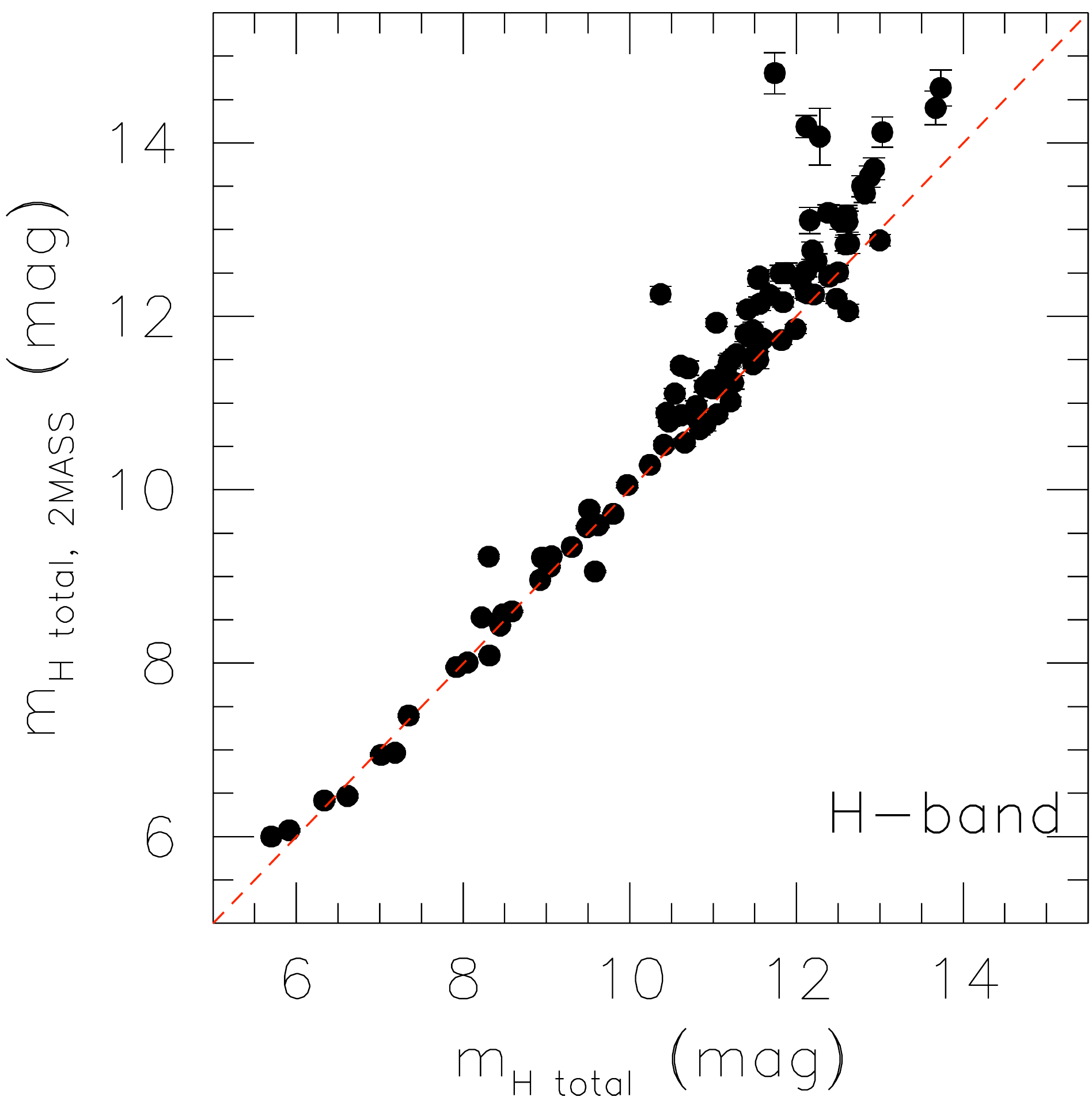} \\
\end{tabular}
\begin{tabular}{cc}
\includegraphics[width=0.45\textwidth]{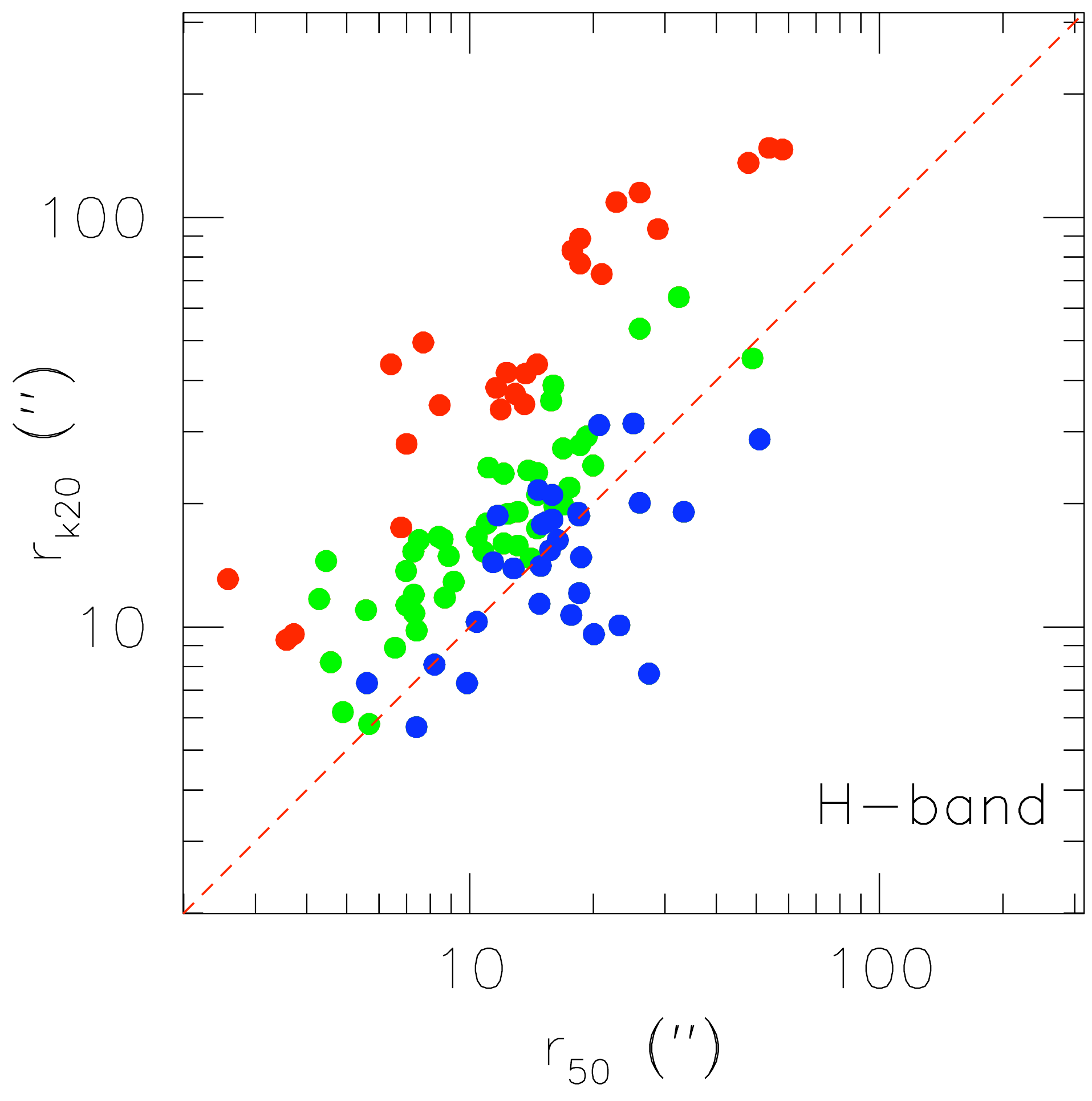} &
\includegraphics[width=0.45\textwidth]{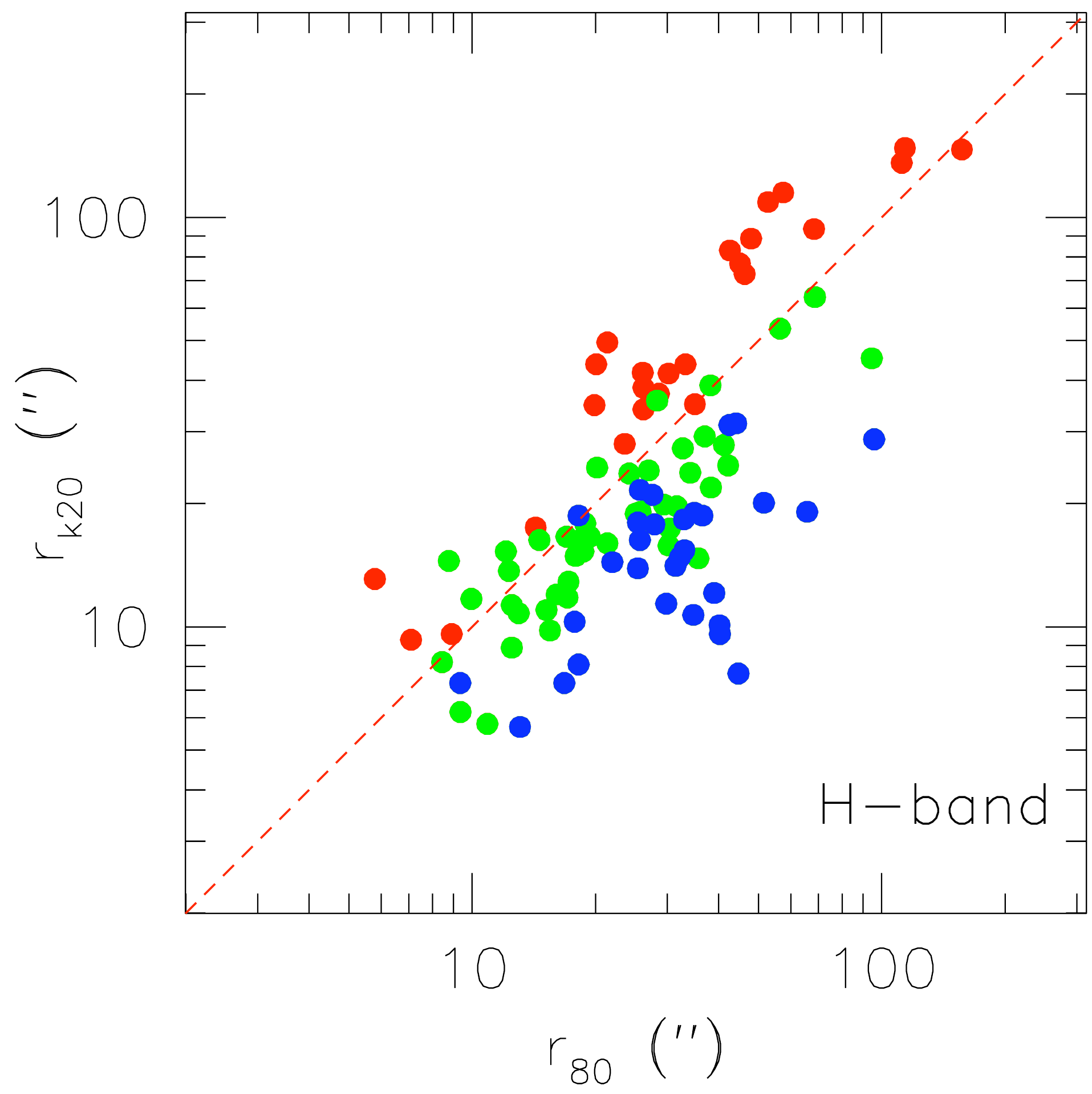} \\
\end{tabular}
\caption{(Top) Comparison of the 2MASS total $H$-band magnitude
with our total isophotal $H$-band magnitude.  Up to 12 $H$-mag,
the agreement is superb with a typical 1-$\sigma$ deviation of 0.2 mag.
The shallower 2MASS is biased towards fainter galaxy magnitudes
for $m_H>12$ mag. (Bottom) Comparison of the 2MASS $K$-band
isophotal radius, $r_{K20}$, measured at the $K$-band surface
brightness of 20 mag arcsec$^{-2}$, with our own 
radii, $r_{H50}$ and $r_{H80}$, which enclose 50\% and 80\%
of the total $H$-band light, respectively.  Modulo the different band-passes, 
the scatter is largely due to the different light
profile shapes for galaxies with different mean surface
brightness.  The red points for $\mu_H <16$ \magarc, the blue 
points for $\mu_H > 18$ \magarc, and green points for the
galaxies in between, make this conclusion very clear. 
The scatter in the distributions of $r_{K20}$, $r_{H50}$ or $r_{H80}$, 
and surface brightness naturally increases for fainter galaxies.}
\label{2ms_compare}
\end{figure*}

We can also compare our well-tested, non-parametric data products,
such as the half-light radius and total magnitude,
with those provided by automated 2MASS and SDSS pipelines.
The upper panel of \Fig{2ms_compare} shows our total, extrapolated
$H$-band brightnesses versus those provided by 2MASS. The 
agreement between the two samples for $H < 12$ mag is excellent, however
the shallower 2MASS catalog is biased for fainter magnitudes.
The lower panels of \Fig{2ms_compare} show our $H$-band radii
against the 2MASS $K$-band isophotal radius, $r_{K20}$, measured
at the $K$-band surface brightness level of 20 mag arcsec$^{-2}$.
We compare $r_{K20}$ to two radii, $r_{50}$ and $r_{80}$,
which enclose 50\% and 80\% of the total $H$-band light,
respectively.  Besides the bandpass differences, the scatter
in this distribution is largely due to the bias between an
absolute metric, $r_{K20}$, and a relative metric, $r_{50}$ or $r_{80}$; 
the former is fixed for any galaxy profile whereas the latter
shifts radially as a function of galaxy profile shape. 
The two measures thus differ as a function of galaxy mean 
surface brightness, as shown in \Fig{2ms_compare}; the agreement
worsens for $\mu_H > 18$ \magarc.  Overall, 
for $m_H < 13$ mag and $\mu_H < 18$ \magarc, the 2MASS total
magnitudes and isophotal radii are reliable to within $\sim$ 0.1 mag and 0.2 dex, respectively.

\begin{figure*}[htb]
\centering
\begin{tabular} {c c}
\includegraphics[width=0.44\textwidth]{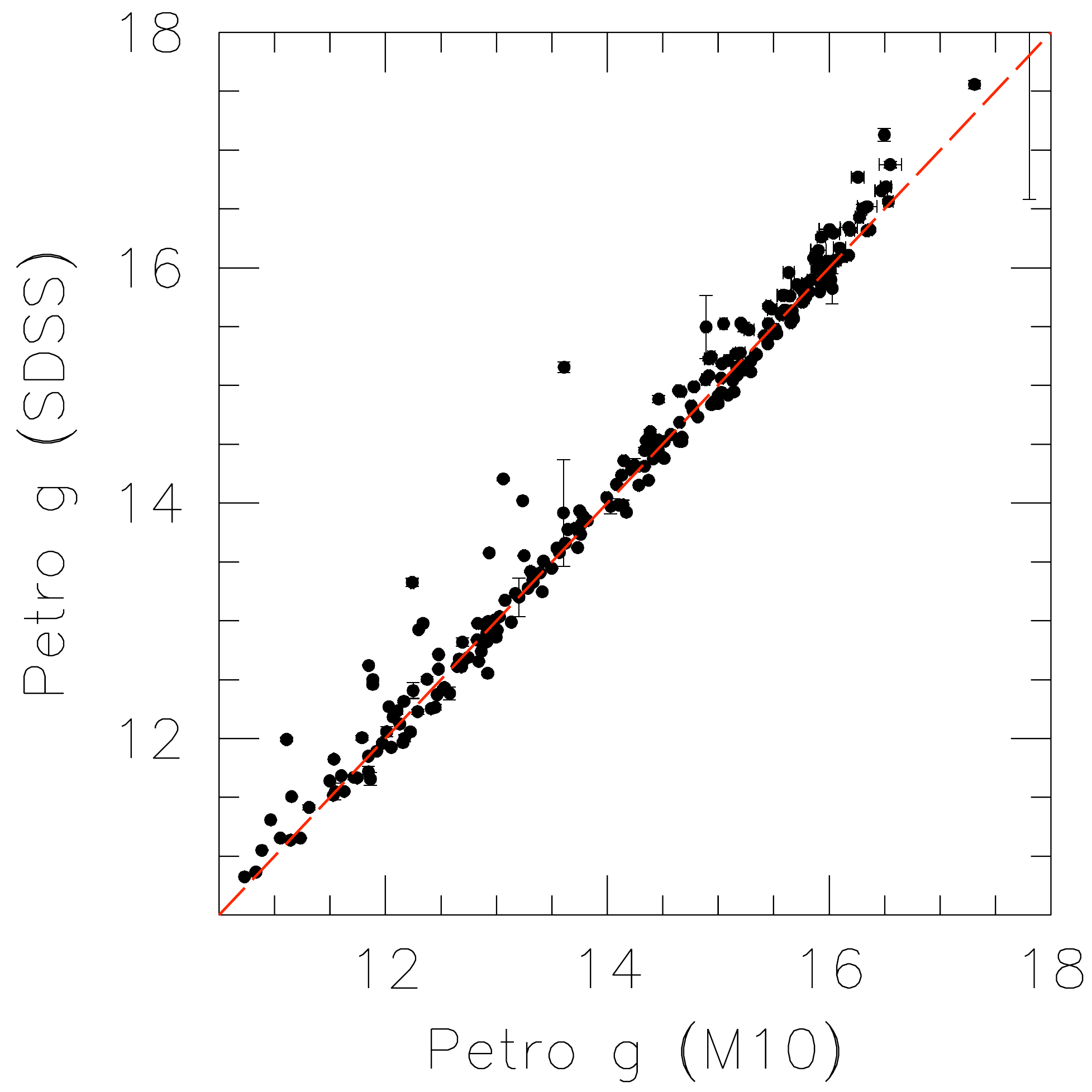} &
\includegraphics[width=0.44\textwidth]{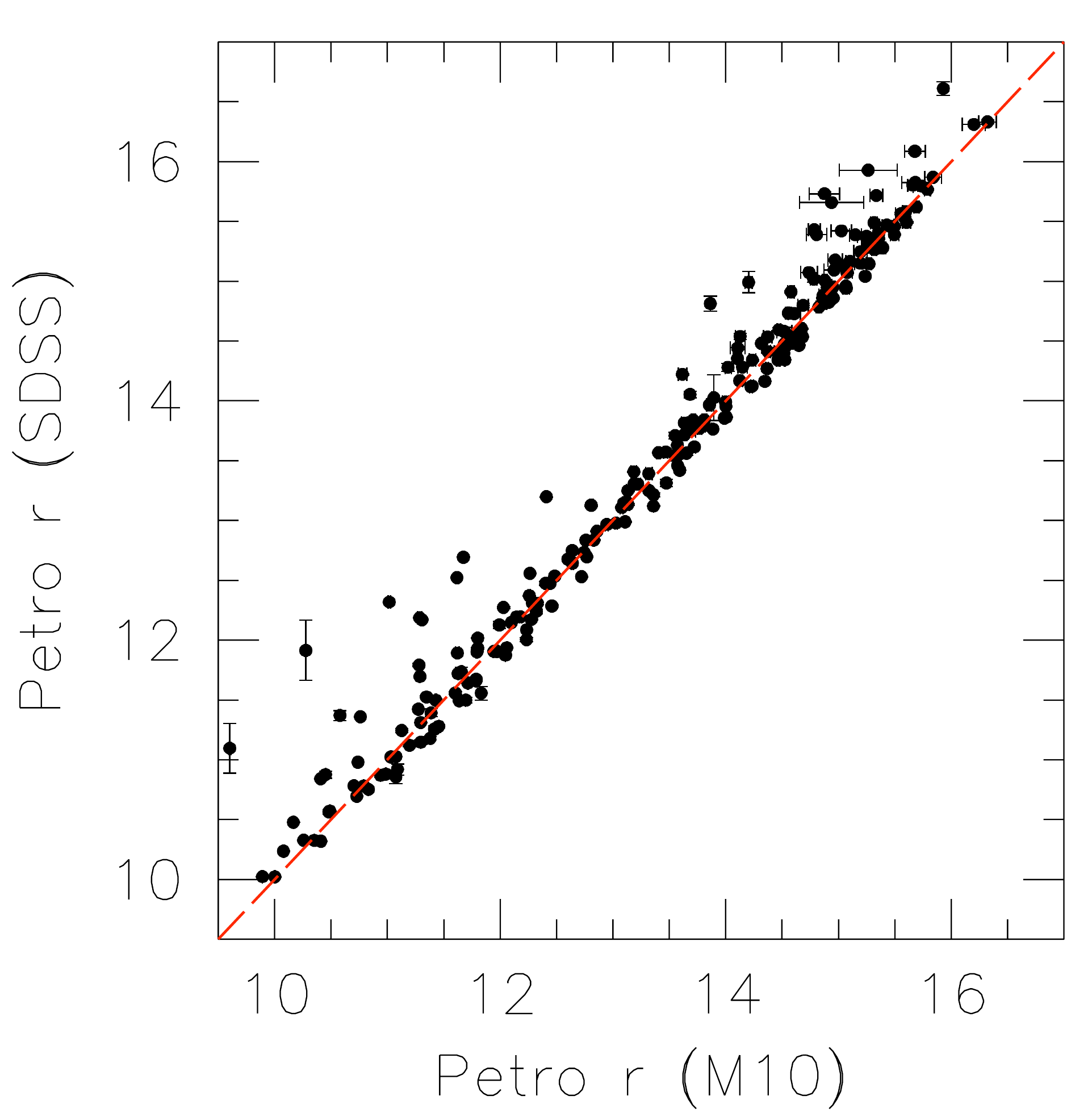} \\
\includegraphics[width=0.44\textwidth]{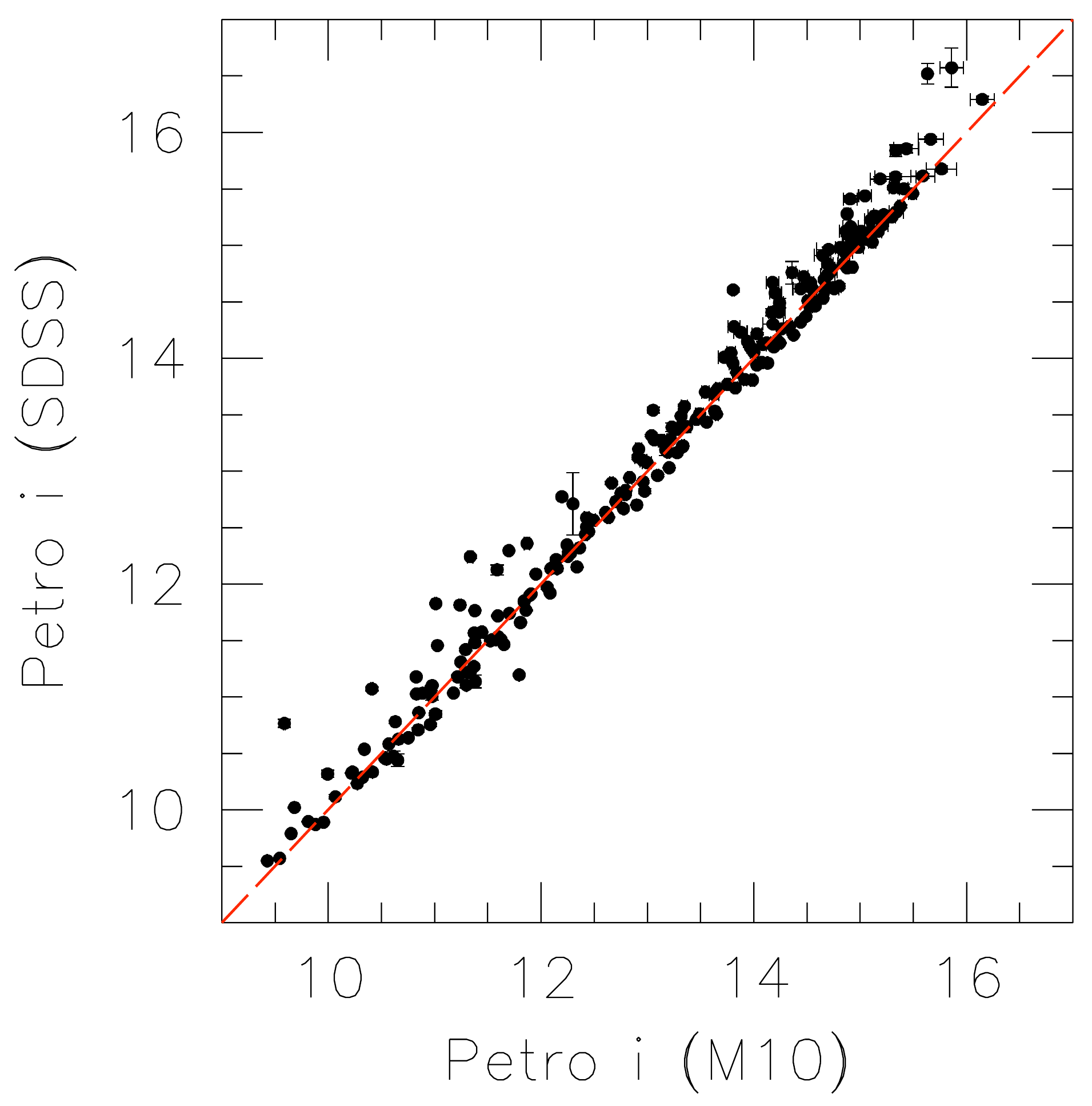} &
\includegraphics[width=0.44\textwidth]{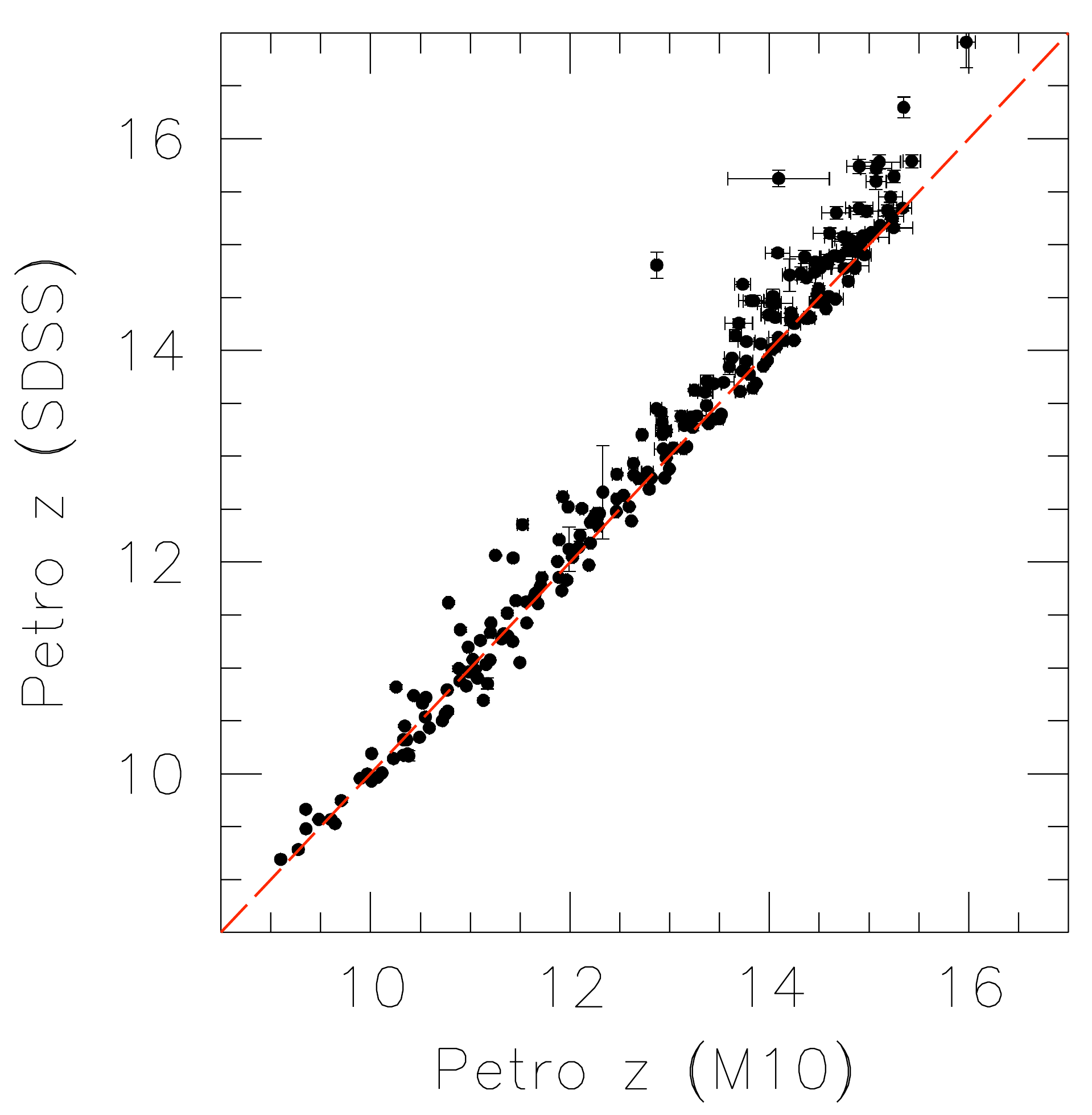} \\
\end{tabular}
\caption{Comparison of available Petrosian magnitudes from the SDSS pipeline and those calculated based on our surface brightness profiles extracted from SDSS images. The red, dashed lines represent the one-to-one relations. These plots confirm that our zeropoint calibration and magnitudes are properly computed.}
\label{sdss_compare1}
\end{figure*}

Before performing a similar comparison with the SDSS data, we
must first confirm that we have properly extracted and calibrated these data.  To that end, we
extract Petrosian magnitudes from our calibrated profiles and
compare these to the published SDSS Petrosian magnitudes (Adelman-McCarthy et al. 2008).  For a 
fair comparison, we use the distributed Petrosian radii and 
computed the total brightness within a circular aperture of
that radius. The results from this analysis, shown in
\Fig{sdss_compare1}, confirm that we reproduce the published
Petrosian magnitudes to within their error for nearly all
galaxies.  This confirms our proper calibration of the SDSS
surface brightness profiles.

\begin{figure*}[htb]
\centering
\begin{tabular} {c}
\includegraphics[width=0.45\textwidth]{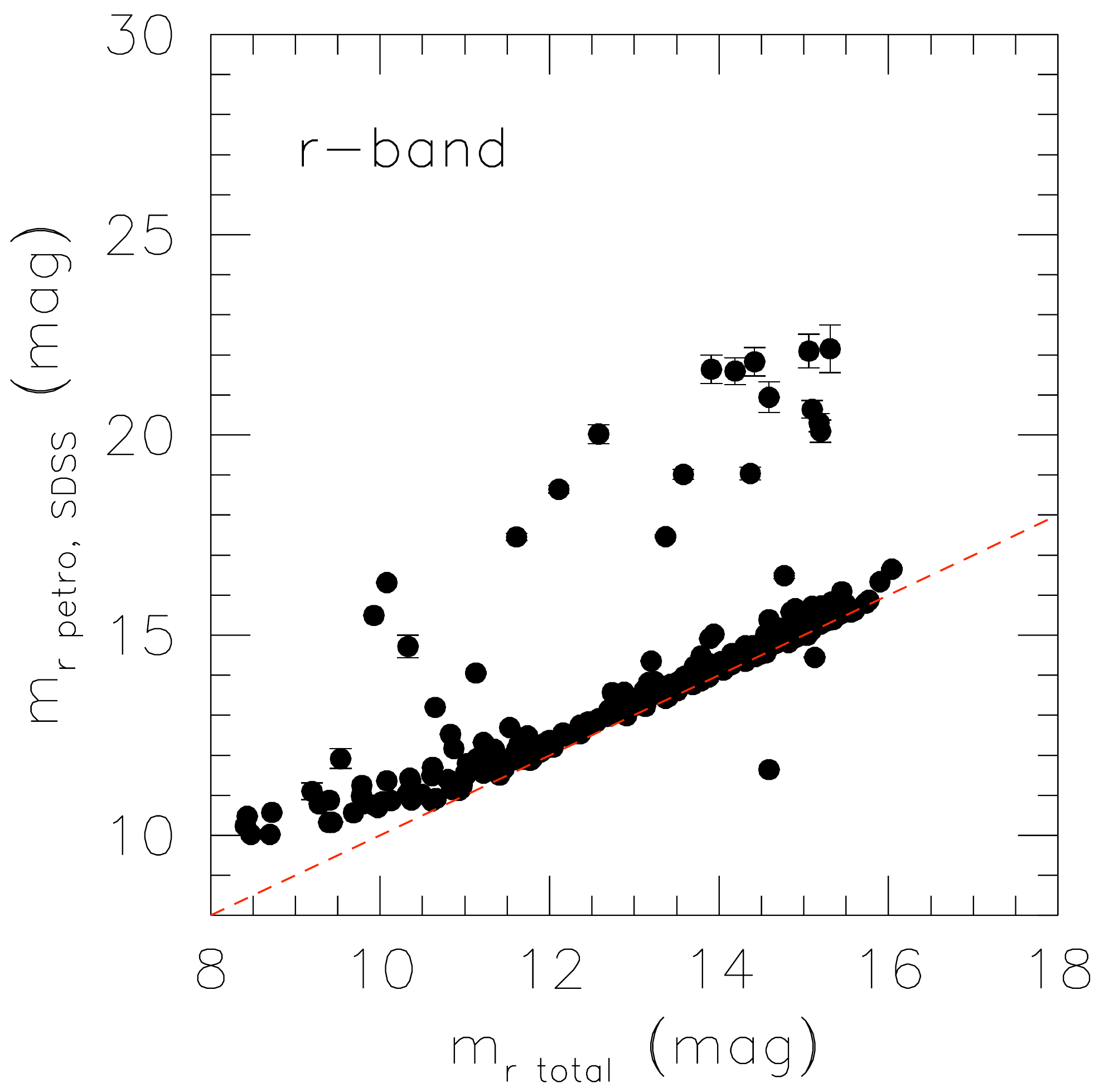} \\
\end{tabular}
\begin{tabular}{cc}
\includegraphics[width=0.45\textwidth]{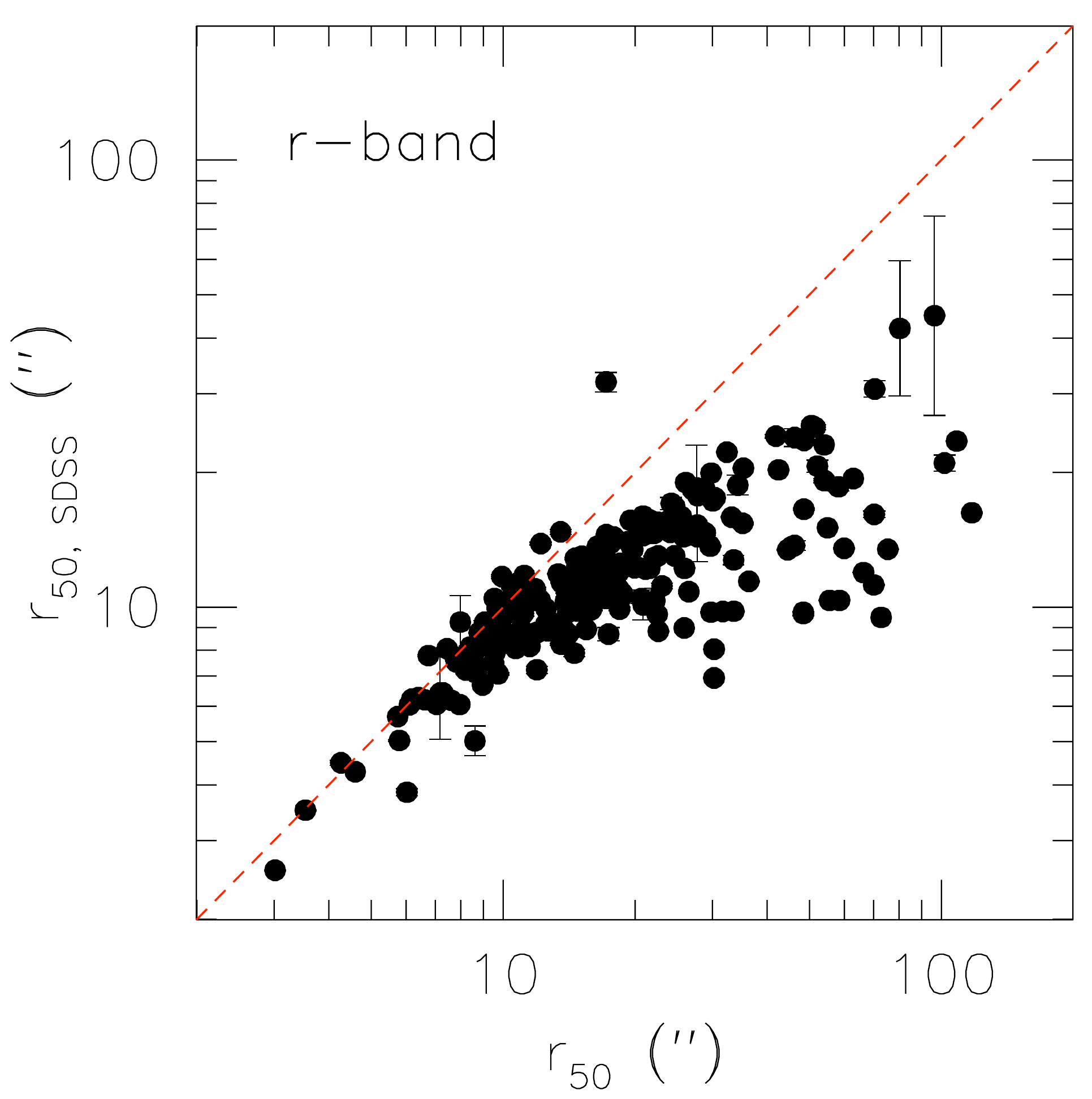} &
\includegraphics[width=0.45\textwidth]{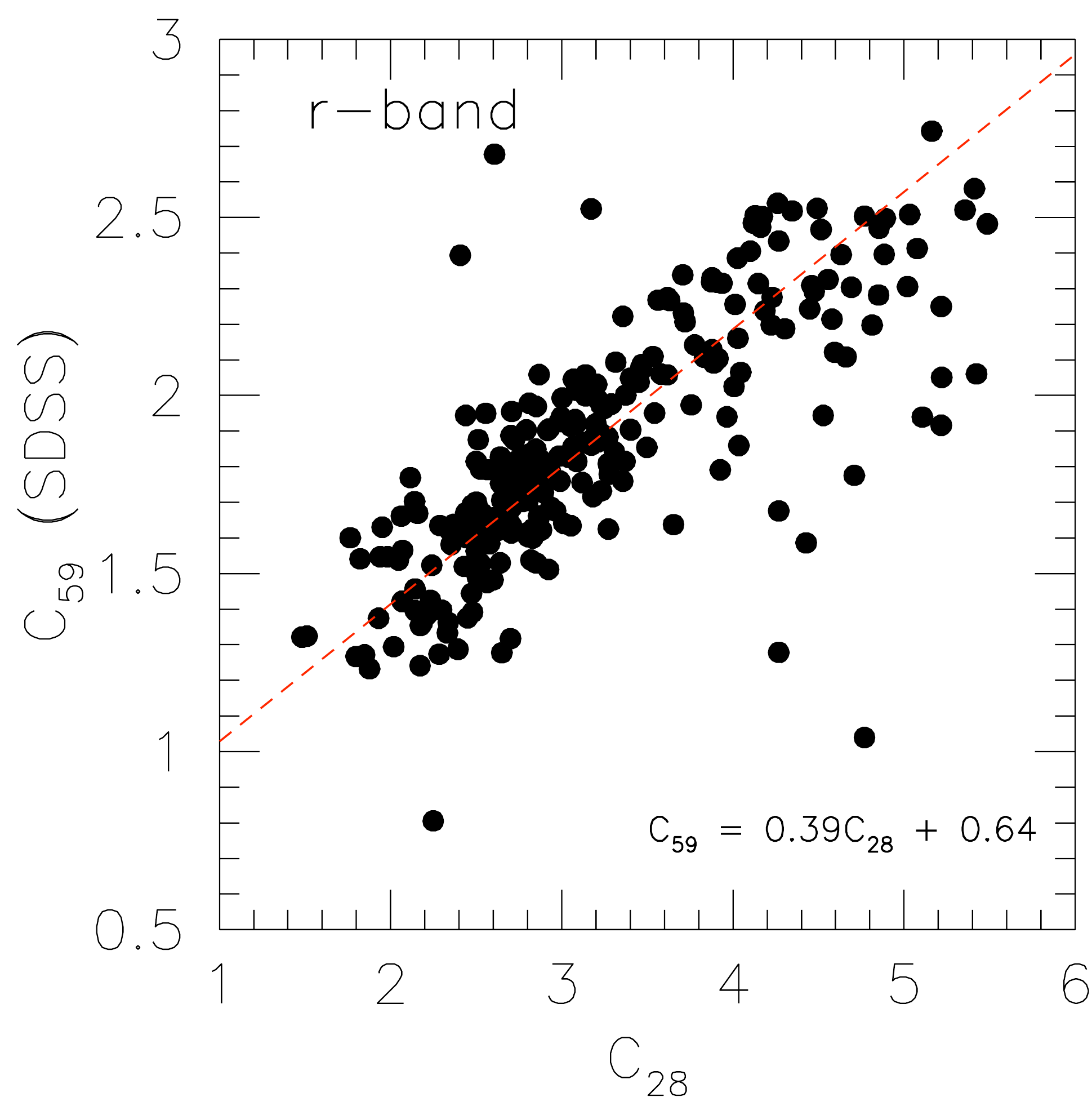} \\
\end{tabular}
\caption{(Top) Comparison of the SDSS Petrosian $r$-band magnitudes
with our total isophotal $r$-band magnitudes from SDSS images.  The
offset at the bright end is due to the SDSS pipeline ``shredding''
large galaxies.  The large scatter is also due to SDSS's 
misidentification of clumpy, star-forming galaxies into multiple sources. 
(Bottom) Left: Comparison of the SDSS Petrosian $r$-band half-light radii,
$r_{50,\rm{SDSS}}$, with the $r$-band half-light radii, $r_{50}$, 
measured with our own software from SDSS images.  The deviation
from the one-to-one line is largely due to comparing radii derived 
from circular (SDSS) apertures versus elliptical apertures.  
Right: Comparison of SDSS concentration, C$_{59}$, with 
our own measure, C$_{28}$.}
\label{sdss_compare2}
\end{figure*}

A comparison with various SDSS data products is shown in
\Fig{sdss_compare2}. The top figure shows the SDSS Petrosian
$r$-band magnitude versus our total isophotal $r$-band
magnitude from SDSS images.  Many galaxies have Petrosian
magnitudes more than 2 magnitudes fainter than ours, likely
due to a galaxy misidentification by the SDSS pipeline
(see Hall \etal 2011, in prep.).   The mean
$\sim$ 0.2 mag offset from the 1-to-1 line for well-behaved
systems is simply the reflection of comparing our total 
magnitudes extrapolated to r~=~$\infty$ with Petrosian
magnitudes which are measured within a finite aperture.
The deviation at the bright end is due to the SDSS
pipeline ``shredding'' of large galaxies (e.g., Panter \etal 2004).  The large scatter at all magnitudes
is most likely due to the misidentification of clumpy, star-forming galaxies into several individual sources.
The bottom left of \Fig{sdss_compare2} shows 
the SDSS Petrosian half-light radius, $r_{50,\rm{SDSS}}$,
against the half-light radius, $r_{50}$, measured with
our own software from SDSS images.  The deviation from
the 1-to-1 line is largely due to comparing radii derived  
from circular (SDSS) apertures versus elliptical apertures, since, at a given radius, a circular aperture will always enclose more flux than an elliptical aperture of the same maximal size.

Finally, the lower-right corner of \Fig{sdss_compare2} shows 
a comparison of our and the SDSS concentration parameter, 
defined as C$_{XY}$=5log$_{10}$($r_{Y}$/$r_{X}$), where 
$r_X$ and $r_Y$ contain $X\times10$\% and $Y\times10$\%
of the total light of the galaxy. 
The curvature between the radii seen in the lower-left panel of \Fig{sdss_compare2}
cancels out to yield a rather linear, albeit noisy, mapping
between the two concentrations.  The conversion between the
two concentration parameters is given by C$_{59}$ = 0.34C$_{28}$ + 0.71. 

Overall, these comparisons highlight the risks of taking survey data at face value (see also Hall \etal 2010).

\section{Surface Brightness Profiles}
\begin{figure*}[htb]
\centering
\includegraphics[width=0.95\textwidth]{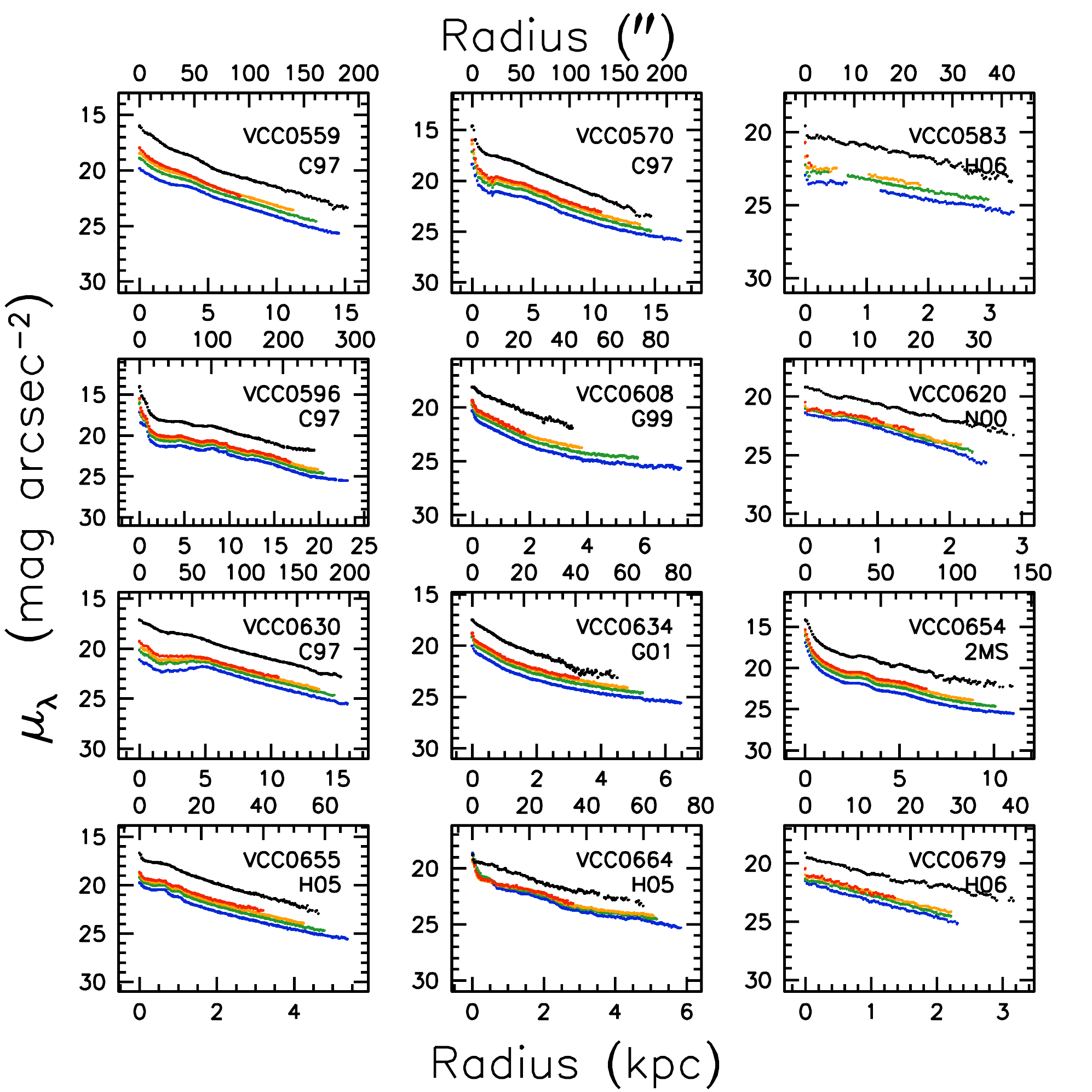}
\caption{Surface brightness profiles for a subset of the 286 galaxies in our sample. The colour coding is as follows: $g$ (blue), $r$ (green), $i$ (orange), $z$ (red), $H$ (black). The VCC galaxy name is included in the upper right corner of each window, with the H-band data source beneath it. The data source codes are: H\#\# (UH 2.2-m ULBCam observations from 2005/06/07/08), CFH (CFHT WIRCAM), UKT (UKIRT WFCAM), 2MS (2MASS archive). All other codes (C\#\#, N\#\#, T\#\#, G\#\#) refer to data taken from the GOLDMine survey and correspond to the telescope used and year of the observation. (The SB profiles for all 286 galaxies are available online at http://www.astro.queensu.ca/virgo/).}
\label{bigfig}
\end{figure*}

\begin{figure*}[htb]
\centering
\includegraphics[width=0.9\textwidth]{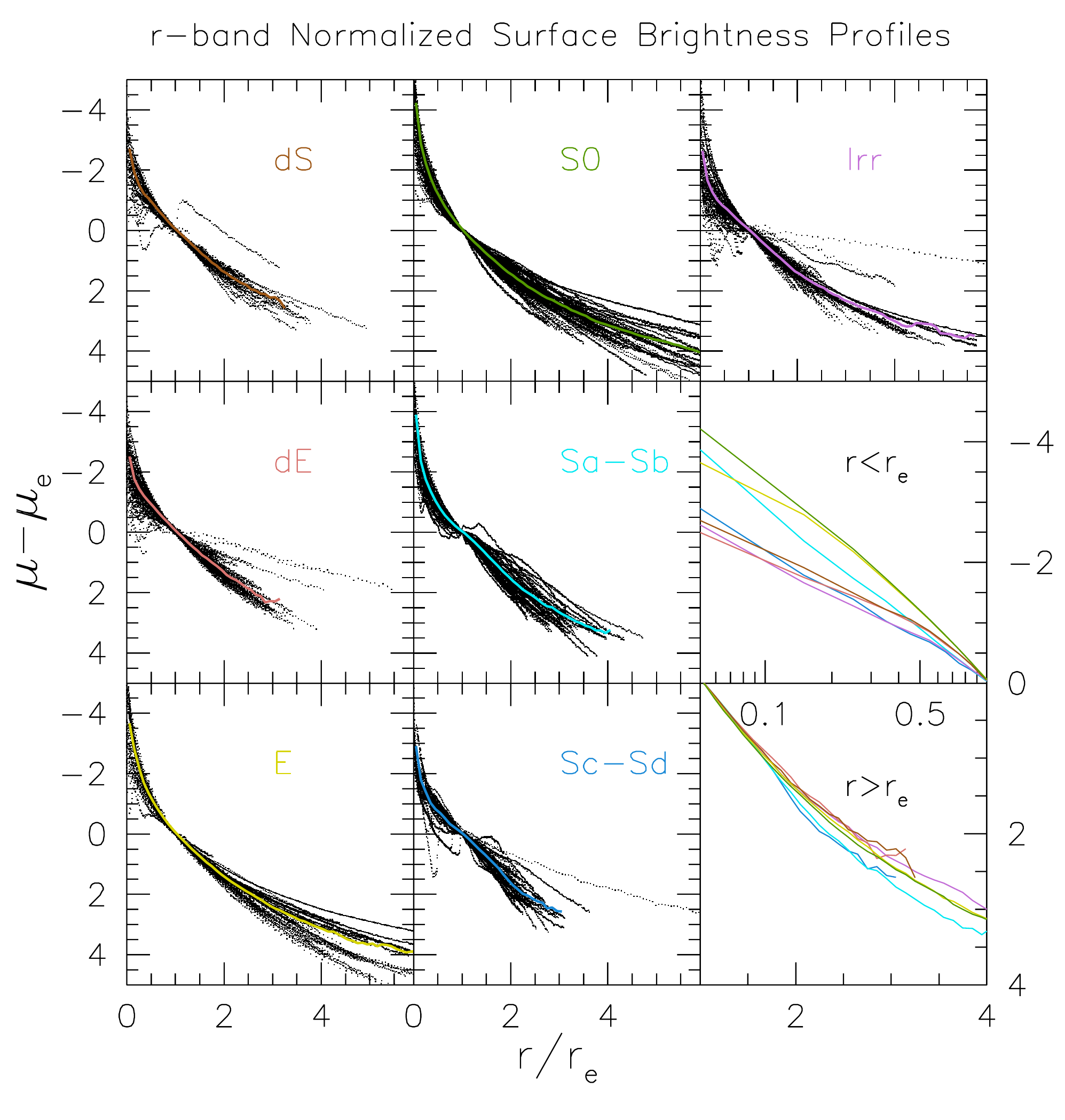}
\caption{$r$-band surface brightness profiles for our entire sample,
  rescaled in terms of $\mu_e$ and $r_e$, and binned by morphological
  classes. Average profiles for each bin are represented by colored
  lines. The middle and lower windows in the right column show the
  average profiles interior and exterior to $r_e$.}
\label{allprofs}
\end{figure*}

We show in \Fig{bigfig} the $griz$ and $H$-band surface
brightness profiles for 12 of the 286 VCC galaxies in our
main sample (see {{\tt http://www.astro.queensu.ca/virgo/}} for the full collection).
The full morphological range spanned by our sample from blue to red and dwarf
to giant is demonstrated by the wide variety of light profile
shapes.  To demonstrate this, we have computed 
the average $r$-band surface brightness profile for galaxies in several
morphological bins.
\Fig{allprofs} shows the differences and similarities between light profiles
at all morphological types, normalized at $\mu_e$ and $r_e$. 
In a plot of $\mu-\mu_e$ versus $r$/$r_e$, all SB profiles must pass
through the point (1,0).  In each window we show the normalised surface
brightness profiles of a given morphology, with the mean profile per
morphological bin shown in color. The middle and lower
windows to the right show the average surface brightness profiles
in each morphological bin interior and exterior to $r_e$.  For $r
\lesssim 0.3r_e$, gas-poor (E, S0, Sa) galaxies exhibit
classical cuspy bulges, while gas-rich types (Sb, Sc, Sd,
dwarf and irregular galaxies) have cored profiles, commonly associated with
pseudo-bulges (MacArthur \etal 2003; Kormendy \& Kennicutt 2004).
For $r \gtrsim 0.3r_e$, the relative shape of the outer profiles
can be compared. Both giant and dwarf spheroidal systems
(E, dE, S0, dS) have similar shapes, while disk-like systems 
(Sa-Sd) exhibit a more exponential outer profile.

\begin{figure*}[htb]
\centering
\includegraphics[width=0.9\textwidth]{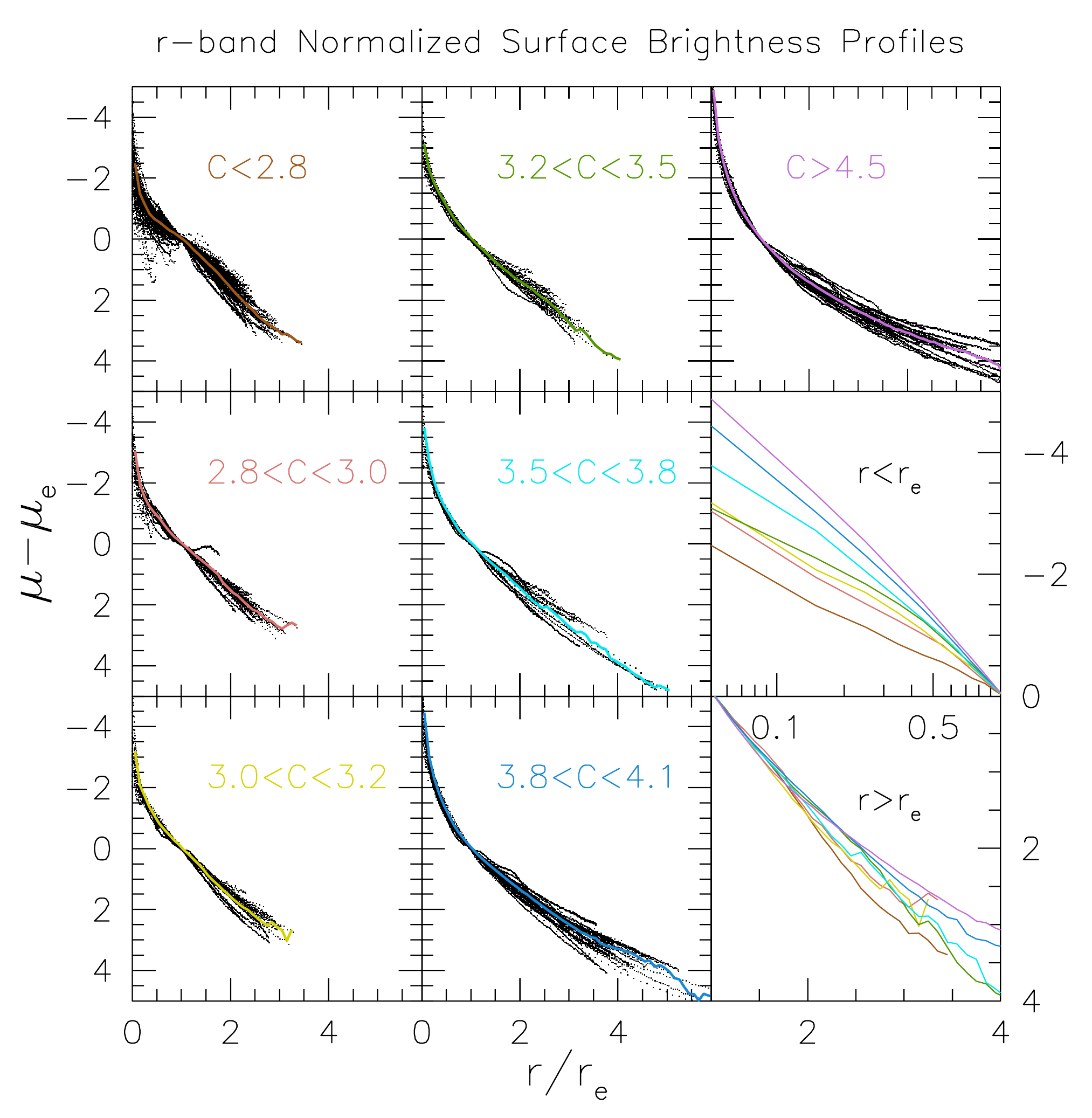}
\caption{$r$-band surface brightness profiles for our entire sample,
  rescaled in terms of $\mu_e$ and $r_e$, and binned by
  concentration. Average profiles for each bin are represented by
  colored lines. The middle and lower windows in the right column
  show the average profiles interior and exterior to $r_e$.}
\label{allprofs_byc}
\end{figure*}

Similar conclusions are reached if we bin the profiles by
concentration rather than morphological type (\Fig{allprofs_byc}).
The use of a quantitive measure for the concentration yields
tighter distributions as opposed to the more subjective
morphological classification. 
There is a natural division for systems with pseudo bulges (C$<$3.8)
versus those with a classical bulge (C$>$3.8), as shown by McDonald \etal (2009b).

We find the same results with the H-band data, though the $r$-band data are, typically, deeper than at H-band. The most noticeable difference is with the depth of the dwarf and
irregular galaxy profiles. The average dE, dS and Irr profiles
are well defined out to much larger radii at $r$-band than at $H$-band. Overall, the trends
that are present in the mean $H$-band profiles are reproduced
at $r$-band, suggesting that these mean profiles are robust. 

Whereas Lauer \etal (2007) hinted at a bimodal separation
of the central regions of giant and dwarf ellipticals into cuspy 
and cored profiles, respectively, C\^{o}t\'{e} \etal 
(2007) showed that the distribution of central profile shapes
is a continuum.  C\^{o}t\'{e} \etal suggested that a possible
selection bias and choice of radius at which the inner slope
is measured could explain the differences, showing
that a continuum of inner slopes is seen in the comparison of the inner profile slope with a non-parametric parameter such as
total luminosity.  Our \Fig{allprofs_byc} attests to the 
continuous distribution of inner slopes when we bin by galaxy
concentration (which correlates with total luminosity). 
Thus, a dichotomy between cuspy and cored
profiles is only seen if profiles are binned by morphology; a continuous 
distribution prevails with a more physically motivated
binning criterion such as concentration. 

\section{Galaxy Structural Parameters}

In order to determine the contribution of the bulge and 
disk components to the total galaxy light, we have performed
1-D bulge-disk decompositions on the full sample of Virgo cluster galaxies following the
techniques described in M09a. Due to the 1-D nature of these decompositions,
we are unable to fully account for non-axisymmetric components such as bars and ovals.
These components will necessarily be added to either the bulge or the disk light, adding
a degree of uncertainty to these fits. Ideally, one should use a 2-D decomposition code such as GIM2D (Simard \etal 2002), BUDDA (de Souza \etal 2004) or GALFIT (Peng \etal 2010) so that non-axisymmetric features can be properly modeled. For further discussion of the relative merits of 1-D and 2-D bulge-disk decompositions, see MacArthur \etal (2003). While 2D decompositions can reduce some of the inherent uncertainty in 1D model parameters, differences between 1D and 2D values are at the 10--20\% level. We proceed with our 1D approach here; our 2D decompositions will be presented elsewhere. We describe the salient features of this method below.

Our bulge model uses a generalized S\'{e}rsic function (S\'{e}rsic 1968)
with three free parameters: an effective (or half-light) radius, $r_e$,
the surface brightness at that radius, $I_e$, and a shape parameter, $n$.
For the special case of $n=1$, the S\'{e}rsic function reduces to
a simple exponential. 
The disk light is modeled with an exponential function described
by two free parameters: a disk central surface brightness, $I_0$, 
and a disk scale length, $h$.  The bulge and disk models are
convolved with the Moffat (1969) function to simulate the effects
of seeing on the intrinsic light profile of a galaxy.

For spiral and irregular galaxies (S0--Sd, Irr) we model the surface brightness profile with a combination of a S\'{e}rsic bulge and exponential disk. If the inclusion of the bulge component does not sufficiently improve the fit quality (i.e. Sd or Irr galaxies), the software will discard it and fit an exponential function alone in an attempt to decrease the degrees of freedom and, thusly, minimize the reduced $\chi^2$. Since spheroidal galaxies (E, dE, dS) may or may not have a disk, we also attempt to model the profile with a single S\'{e}rsic. Once again, if the absence of a disk component increases the goodness-of-fit, the software will automatically discard this component of the model. The fit minimization uses brightnesses in magnitude (not counts), and the lowest global value of the $\chi^2$ per degree of freedom determines the best fit. 

Our model decomposition can also account for the presence
of a nucleus, here modeled by a seeing-convolved delta function at $r=0$ (one additional free
parameter, $m_{nuc}$), as well as spiral arms, which are
modeled by a smooth increase in brightness above the
underlying disk. We find that the addition of a nuclear component reduces the need for high-$n$ bulges, yielding an average shape parameter of $n\sim$ 1. These additional features increase the
degrees of freedom and, thus, are only applied for obvious
cases. 

Non-parametric structural measurements, such as the total
magnitude (extrapolated from the last data point to infinity),
the effective radius and surface brightness, and galaxy light concentration, were computed for all 286 galaxies in each of the 5 bands. These measurements are nearly independent of the bulge-disk model decompositions, relying only on the fit in order to determine the amount of galaxy light beyond the last data point (typically $<$ 0.2 mag). The fact that these parameters are independent of any model (i.e. no assumption that a galaxy profile can be separated into components) makes them most useful for galaxy structure studies. The effective radius,
$r_e$, is defined as the radius which encloses half the total light, 
while the effective surface brightness is the surface brightness at this radius, $\mu_e=\mu(r_e)$. 
We compute the concentration parameter, C$_{28}$, as in \se{other}. 

\begin{figure*}[htb]
\centering
\includegraphics[width=0.9\textwidth]{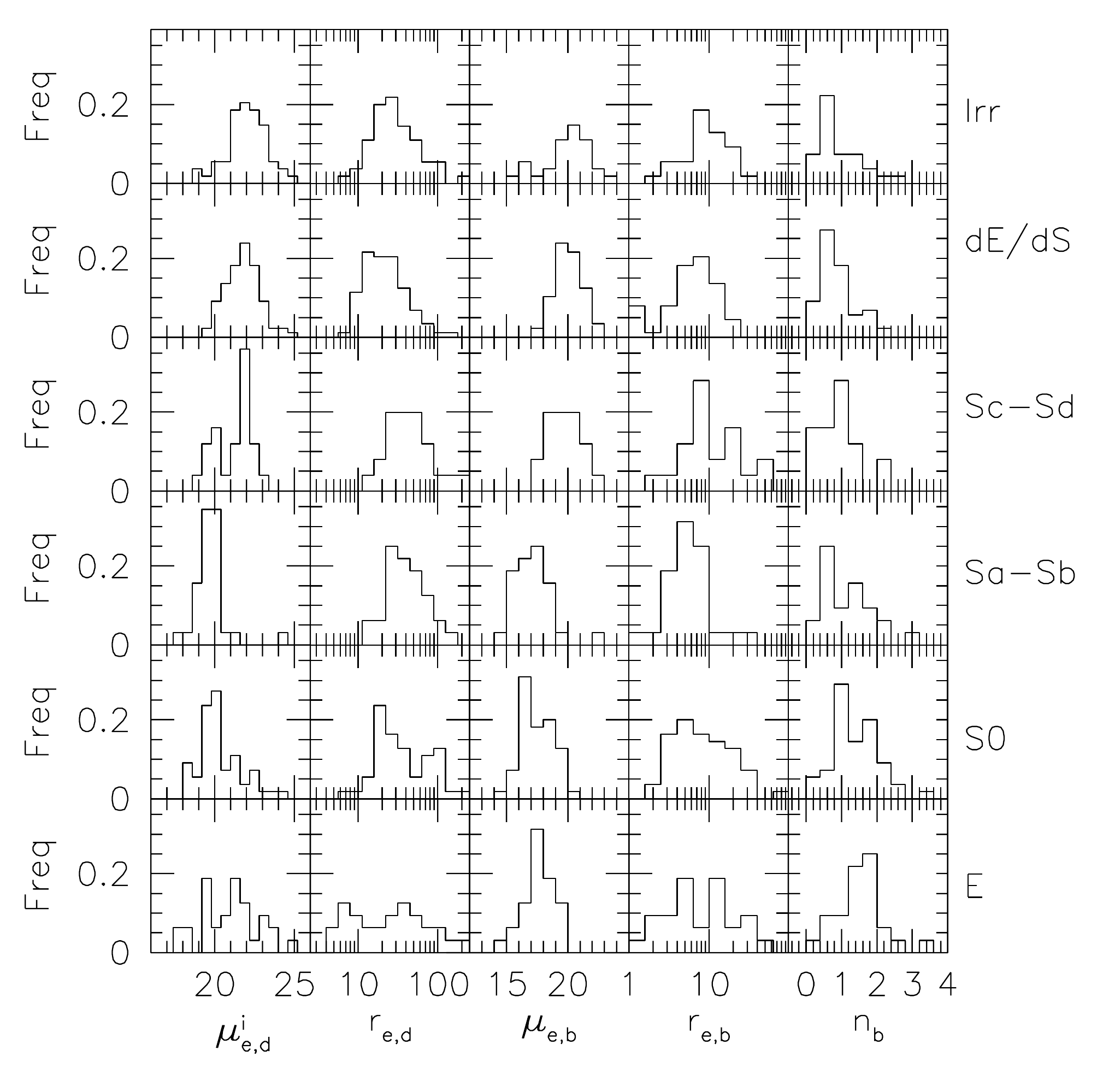}
\caption{Distribution of H-band structural parameters from bulge-disk decompositions for Virgo galaxies sorted into 6 different morphological bins.}
\label{avgpars}
\end{figure*}

The parametric and non-parametric structural measurements
for our 286 VCC galaxies in the $griz$ and $H$ bands are provided online at the CDS ({\tt http://cds.u-strasbg.fr/}) and our private website ({{\tt http://www.astro.queensu.ca/virgo/}}). For spiral galaxies with well-defined bulges and disks (i.e., Sa-Sb), we find typical $r$-band errors on our fitting parameters of: $\Delta\mu_{e,d}=0.1$ \magarc, $\Delta\mu_{e,b}=0.2$ \magarc, $\Delta r_{e,d}=3$\%, $\Delta r_{e,b}=15$\%, and $\Delta n_b=0.3$. These uncertainties are similar to those reported by MacArthur \etal (2003), who relied on similar techniques. For irregular or poorly-sampled systems, these uncertainties can increase by as much as an order of magnitude. The distribution of the H-band parametric quantities
is shown in \Fig{avgpars} for different morphological bins. This figure clearly shows the disk surface brightness bimodality discovered by TV97 and confirmed by McDonald \etal (2009a,b). These
distributions are meant to be compared to 
galaxy formation models. 

\section{Tables of Median Structural Parameters}

We provide in Table 1 the median and standard deviation
of the four non-parametric parameters (total magnitude, effective
surface brightness, effective radius, and concentration) as a
function of morphology and bandpass.  This table shows the variation
of galaxy structural parameters with wavelength and morphology, as well as the
spread of each parameter within a given morphological/wavelength bin. 
For instance, there is a substantial range in elliptical
galaxy brightnesses, as expected from the bimodal distribution
of elliptical surface brightnesses (McDonald \etal 2009b), and a 
clear division of galaxies into cuspy 
(C$_{28} \sim 4.0$) and disk-like (C$_{28} \sim 2.8$)
galaxies, also as reported by McDonald \etal (2009b).
Additionally, Table 1 shows that disk galaxies have, on average,
effective radii that are twice as large as spheroidal galaxies,
regardless of surface brightness or color. 

Table 2 gives the median and standard deviation of five parametric
quantities from 1D bulge-disk decompositions. This includes
three bulge parameters: $\mu_{e,b}$, $r_{e,b}$, $n$ and the
two disk parameters: $\mu_{e,d}$ and $r_{e,d}$.  This table
shows the variation of bulge and disk structural parameters
as a function of color and morphology.  We find that the 
``typical'' bulge shape is indeed exponential (e.g., Courteau
\etal 1996; MacArthur \etal 2003) and that the light profile
of a typical elliptical galaxy is well fit by a single S\'{e}rsic
index of $n\sim2$. No Virgo galaxy spheroid requires $n > 3$. 
Futhermore, the decomposition of elliptical galaxy
light profiles requires the addition of a disk component for
$\sim$ 70\% of our sample.  These disks tend to have low surface brightness, with average brightnesses similar to Sc-Sd galaxies, but typically have scale lengths twice as large. Finally, we find that the typical spiral galaxy disk has a 
half-light radius $\sim 5\times$ larger than that of the bulge.

Tables 1 and 2 offer a wealth of information about the structural parameters of galaxies across the Hubble sequence and over a range of wavelengths. An extensive discussion of these trends is found in McDonald \etal (2009b).  These tables ought to provide a valuable compendium of galaxy structural properties for theoretical investigations of galaxies
as well as a much needed foil for comparisons with high redshift galaxy samples. 

\section{Summary}
We have presented the results of a NIR survey, supplemented with optical imaging from the SDSS, of 286 morphologically-diverse Virgo cluster galaxies. Our NIR images come from various sources including archival $H$-band data from the 2MASS and GOLDMine archives and new $H$-band data for 171 galaxies from the UH 2.2-m, CFHT and UKIRT telescopes. These NIR data have been carefully calibrated, yielding a photometric accuracy of $\sim$ 0.1 mag arcsec$^{-2}$ over the entire sample. Both the new and archival data were analyzed in a homogeneous manner using our own software
for uniformity. For each galaxy in our sample we have extracted $g$,$r$,$i$,$z$,$H$ surface brightness profiles yielding a deep, multiwavelength collection of surface photometry. Full bulge-disk
decompositions were performed in each of the five wavebands
for all 286 VCC galaxies. We also provide the average structural parameters for galaxies in various morphological bins, offering a most valuable benchmark for comparison against theoretical investigations of galaxies as well as high redshift galaxy samples. This database has recently been utilized in studies of galaxy structural parameters by McDonald \etal (2009b) and stellar populations by Roediger \etal (2010) and Prochaska \etal (2011). The galaxy light profiles and results of bulge-disk decompositions are available online at the CDS ({\tt http://cds.u-strasbg.fr/}) and our own website ({{\tt http://www.astro.queensu.ca/virgo/}}).

\section*{Acknowledgements}

S.C. acknowledges support through 
a Discovery Grant of the Natural Science
and Engineering Research Council of Canada.
R. B. T. acknowledges support from US National
Science Foundation award AST 09-08846. 
Guiseppe Gavazzi is thanked for sharing Goldmine data
 products with us.  We are also indebted to Enrico Maria
 Corsini for catching a subtle error in our computation
 of sky errors and enabling its correction before going
 to print.  Lastly, we are grateful to Jon Holtzman and
 Melanie Hall for their contributions to the data selection,
 acquisition and reduction phases of this survey.

\clearpage

\clearpage


\begin{table}[htb]
\label{tableofmeans1}
\caption[]{Median and standard deviation of Virgo cluster galaxy non-parametric structural quantities for different morphologies and bandpasses.}
\begin{center}
\begin{tabular}{c c c c c c c}
\hline\hline
Parameter & Morphology & $g$ & $r$ & $i$ & $z$ & $H$ \\
\hline
\\
m$_T$ & E & 13.5(1.8) & 12.9(1.8) & 12.4(1.8) & 12.7(1.9) & 10.6(2.2)\\
$[$mag$]$ & S0 & 12.5(1.3) & 11.6(1.4) & 11.2(1.4) & 11.0(1.5) & 9.3(1.6)\\
(1) & Sa-Sb & 12.1(1.2) & 11.5(1.3) & 11.1(1.3) & 10.9(1.3) & 9.0(1.5)\\
 & Sc-Sd & 13.3(1.3) & 12.7(1.5) & 12.6(1.4) & 12.4(1.5) & 9.9(1.8)\\
 & dE & 15.3(0.9) & 14.7(0.8) & 14.4(0.9) & 14.2(1.3) & 12.5(0.9)\\
 & dS & 14.7(0.5) & 14.1(0.6) & 13.8(0.6) & 13.6(0.7) & 11.7(0.9)\\
 & Irr & 14.9(1.3) & 14.3(1.4) & 14.0(1.5) & 13.7(1.5) & 12.3(1.6)\\
\\
$\mu_e$ & E & 22.0(1.4) & 21.2(1.5) & 20.9(1.5) & 20.5(1.4) & 18.3(1.5)\\
$[$mag arcsec$^{-2}]$ & S0 & 22.3(0.9) & 21.5(1.0) & 21.1(1.0) & 20.8(1.0) & 18.9(1.1)\\
(2) & Sa-Sb & 22.8(0.7) & 22.2(0.7) & 21.8(0.7) & 21.5(0.7) & 19.5(0.7)\\
 & Sc-Sd & 23.5(0.7) & 23.0(0.7) & 22.7(0.7) & 22.3(0.7) & 20.6(1.0)\\
 & dE & 24.1(0.7) & 23.4(0.7) & 23.1(0.8) & 22.6(0.7) & 21.1(0.9)\\
 & dS & 23.8(0.9) & 23.2(0.8) & 22.9(0.8) & 22.5(0.8) & 20.6(0.8)\\
 & Irr & 23.7(1.0) & 23.3(1.1) & 23.1(1.1) & 22.6(1.1) & 21.3(1.3)\\
\\
$r_e$ & E & 1.1(0.8) & 1.1(0.8) & 1.1(0.9) & 1.0(0.7) & 0.9(1.1)\\
$[$kpc$]$ & S0 & 1.7(0.9) & 1.8(0.9) & 1.7(0.9) & 1.6(0.8) & 1.4(1.0)\\
(3) & Sa-Sb & 2.6(1.6) & 2.5(1.5) & 2.9(1.4) & 2.5(1.4) & 2.5(2.0)\\
 & Sc-Sd & 2.4(1.8) & 2.4(1.7) & 2.4(1.5) & 2.3(1.4) & 2.2(2.4)\\
 & dE & 1.3(0.5) & 1.3(0.5) & 1.2(0.5) & 1.0(0.8) & 1.2(0.5)\\
 & dS & 1.2(0.6) & 1.2(0.5) & 1.2(0.5) & 1.2(0.5) & 1.1(0.5)\\
 & Irr & 1.4(0.7) & 1.4(0.7) & 1.5(0.8) & 1.3(0.7) & 1.3(0.8)\\
\\
C$_{28}$ & E & 3.8(0.7) & 3.9(0.7) & 4.0(0.7) & 3.8(1.0) & 3.8(0.7)\\
(4) & S0 & 4.0(0.8) & 4.1(0.8) & 4.1(0.8) & 4.2(0.8) & 4.0(0.9)\\
 & Sa-Sb & 3.1(0.7) & 3.2(0.7) & 3.2(0.7) & 3.3(0.7) & 3.4(0.7)\\
 & Sc-Sd & 2.6(0.4) & 2.5(0.5) & 2.8(0.5) & 2.7(0.6) & 2.8(0.4)\\
 & dE & 2.9(0.6) & 3.0(0.6) & 2.9(0.6) & 2.7(0.5) & 2.9(0.5)\\
 & dS & 3.3(0.4) & 3.3(0.4) & 3.2(0.4) & 2.9(0.4) & 3.3(0.6)\\
 & Irr & 2.7(0.8) & 2.7(0.8) & 2.6(0.7) & 2.5(0.7) & 2.6(0.7)\\
\\
\hline

\end{tabular}
\end{center}
{\footnotesize
(1) Total apparent magnitude extrapolated to r=$\infty$ (mag)
(2) Effective surface brightness (mag arcsec$^{-2}$)
(3) Effective radius (kpc)
(4) Concentration: C$_{28}$=5log$_{10}$($r_{80}$/$r_{20}$) 
}
\end{table}

\begin{table}[htb]
\label{tableofmeans2}
{\footnotesize
\caption[]{Median and standard deviation of Virgo cluster galaxy parametric structural parameters from bulge-disk decompositions for different morphologies and bandpasses.}
\begin{center}
\begin{tabular}{c c c c c c c}
\hline\hline
Parameter & Morphology & $g$ & $r$ & $i$ & $z$ & $H$ \\
\hline
\\
$n$ & E$^1$ & 2.4(1.1) & 2.6(1.0) & 2.3(1.0) & 2.2(1.4) & 2.1(0.9)\\
(1) & E$^2$ & 1.4(0.6) & 1.5(0.5) & 1.5(0.6) & 1.3(0.7) & 1.5(0.7)\\
 & S0 & 1.4(0.6) & 1.4(0.5) & 1.4(0.6) & 1.3(0.6) & 1.2(0.7)\\
 & Sa-Sb & 1.2(0.7) & 1.2(0.6) & 1.4(0.6) & 1.2(0.7) & 1.0(0.6)\\
 & Sc-Sd & 0.5(0.3) & 0.7(0.5) & 0.6(0.4) & 0.6(0.4) & 0.8(0.5)\\
 & dE/dS & 0.9(0.3) & 0.9(0.4) & 0.9(0.4) & 0.8(0.4) & 0.8(0.5)\\
 & Irr & 0.6(0.6) & 0.8(0.6) & 0.7(0.5) & 0.5(0.5) & 0.7(0.6)\\
\\
$\mu_{e,b}$ & E$^1$ & 22.7(1.4) & 21.7(1.4) & 21.4(1.4) & 21.3(1.5) & 18.7(0.9)\\
$[$mag arcsec$^{-2}]$ & E$^2$ & 21.0(1.4) & 20.5(1.4) & 20.2(1.4) & 19.8(1.6) & 17.5(1.3)\\
(2) & S0 & 20.6(1.3) & 20.0(1.3) & 19.4(1.4) & 19.2(1.5) & 17.2(1.3)\\
 & Sa-Sb & 21.2(1.2) & 20.5(1.2) & 20.3(1.4) & 19.9(1.4) & 17.6(1.5)\\
 & Sc-Sd & 22.6(1.5) & 21.9(0.9) & 21.6(1.0) & 21.4(1.2) & 19.7(1.4)\\
 & dE/dS & 23.0(0.9) & 22.6(0.9) & 22.2(1.0) & 21.9(0.9) & 20.0(1.1)\\
 & Irr & 22.6(1.4) & 22.1(1.5) & 21.8(1.6) & 21.0(1.7) & 20.2(1.8)\\
\\
$r_{e,b}$ & E$^1$ & 1.7(2.3) & 1.6(2.0) & 1.5(2.0) & 1.5(3.2) & 1.3(2.0) \\
$[$kpc$]$ & E$^2$ & 0.6(0.4) & 0.6(0.3) & 0.6(0.3) & 0.6(0.3) & 0.7(0.9) \\
(3) & S0 & 0.8(0.6) & 0.8(0.6) & 0.6(0.7) & 0.6(0.6) & 0.6(0.9) \\
 & Sa-Sb & 0.5(1.4) & 0.6(0.6) & 0.6(0.8) & 0.6(0.7) & 0.5(0.6) \\
 & Sc-Sd & 0.5(0.4) & 0.4(0.4) & 0.6(0.6) & 0.6(0.5) & 0.7(1.1) \\
 & dE/dS & 0.6(0.4) & 0.6(0.3) & 0.6(0.3) & 0.5(0.3) & 0.5(0.4) \\
 & Irr & 0.6(0.4) & 0.8(0.5) & 0.8(0.4) & 0.4(0.3) & 0.8(0.5) \\
\\
$\mu_{e,d}$ & E$^2$ & 23.8(1.6) & 23.3(1.6) & 22.7(1.6) & 22.3(1.6) & 20.9(1.8)\\
$[$mag arcsec$^{-2}]$ & S0 & 22.9(1.5) & 22.3(1.3) & 21.9(1.2) & 21.3(1.4) & 19.6(1.5)\\
(4) & Sa-Sb & 22.6(1.3) & 22.0(0.9) & 21.6(1.1) & 21.3(1.1) & 18.9(1.1)\\
 & Sc-Sd & 22.9(0.8) & 22.4(0.9) & 22.1(1.1) & 21.9(0.8) & 20.5(1.4)\\
 & dE/dS & 24.3(1.0) & 23.7(0.9) & 23.4(0.9) & 22.9(0.9) & 21.4(1.0)\\
 & Irr & 23.8(1.2) & 23.4(1.1) & 23.2(1.1) & 22.6(0.9) & 21.4(1.3)\\
\\
$r_{e,d}$ & E$^2$ & 2.4(2.3) & 2.4(2.6) & 2.4(2.4) & 2.1(2.5) & 2.5(9.1) \\
$[$kpc$]$ & S0 & 3.3(4.0) & 3.4(3.5) & 3.0(5.0) & 2.5(5.1) & 2.9(7.5) \\
(5) & Sa-Sb & 3.7(3.6) & 3.4(2.2) & 3.9(2.0) & 3.3(2.7) & 3.3(2.6) \\
 & Sc-Sd & 2.6(2.7) & 2.7(2.5) & 2.8(2.5) & 2.9(2.1) & 3.0(10.7) \\
 & dE/dS & 1.9(1.8) & 1.9(1.8) & 1.9(1.4) & 1.8(2.3) & 1.8(1.8) \\
 & Irr & 2.2(1.8) & 2.1(2.3) & 2.4(1.6) & 2.0(1.2) & 2.2(3.0) \\
\\
\hline

\end{tabular}
\end{center}
(1) S\'{e}rsic n parameter
(2) Bulge effective surface brightness in units of mag arcsec$^{-2}$
(3) Bulge effective radius in units of kpc
(4) Disk effective surface brightness in units of mag arcsec$^{-2}$
(5) Disk effective radius in units of kpc
(E$^1$) Elliptical galaxy light profile fit with single Sersic function
(E$^2$) Elliptical galaxy light profile fit with Sersic + exponential functions
}
\end{table}

\end{document}